\documentstyle[sprocl]{article}

\input epsf
\newcommand\epjc[3]{{\em Eur.\ Phys.\ J.\ C }{\bf #1}, #3 (#2)}
\newcommand\npb[3]{{\em Nucl.\ Phys.\ B }{\bf #1}, #3 (#2)}
\newcommand\npps[3]{{\em Nucl.\ Phys.\ (Proc.\ Suppl.) B }
                    {\bf #1}, #3 (#2)}
\newcommand\plb[3]{{\em Phys.\ Lett.\ B }{\bf #1}, #3 (#2)}
\newcommand\prd[3]{{\em Phys.\ Rev.\ D }{\bf #1}, #3 (#2)}
\newcommand\prep[3]{{\em Phys.\ Rep.\ }{\bf #1}, #3 (#2)}
\newcommand\prl[3]{{\em Phys.\ Rev.\ Lett.\ }{\bf #1}, #3 (#2)}
\newcommand\rmp[3]{{\em Rev.\ Mod.\ Phys.\ }{\bf #1}, #3 (#2)}
\newcommand\zpc[3]{{\em Z.\ Phys.\ C }{\bf #1}, #3 (#2)}
\newcommand\sjnp[3]{{\em Sov.\ J.\ Nucl.\ Phys.\ }{\bf #1}, #3 (#2)}
\newcommand\yf[3]{{\em Yad.\ Fiz.\ }{\bf #1}, #3 (#2)}
\newcommand\ibid[3]{{\em ibid.\ }{\bf #1}, #3 (#2)}
\newcommand{\hepph}[1]{{\tt hep-ph/#1}}
\newcommand{\hepex}[1]{{\tt hep-ex/#1}}

\begin{document}

\begin{flushright}
CLNS~00/1712\\
{\tt hep-ph/0012204}
\end{flushright}
\vspace{0.5cm}

\title{LECTURES ON THE THEORY OF\\
NON-LEPTONIC B DECAYS}

\author{MATTHIAS NEUBERT}

\address{Newman Laboratory of Nuclear Studies, Cornell University\\
Ithaca, NY 14853, USA\\
E-mail: neubert@mail.lns.cornell.edu} 

\maketitle\abstracts{
These notes provide a pedagogical introduction to the theory of
non-leptonic heavy-meson decays recently proposed by Beneke, Buchalla,
Sachrajda and myself. We provide a rigorous basis for factorization 
for a large class of non-leptonic two-body $B$-meson decays in the 
heavy-quark limit. The resulting factorization formula incorporates 
elements of the naive factorization approach and the hard-scattering 
approach, and allows us to compute systematically radiative 
(``non-factorizable'') corrections to naive factorization for decays 
such as $B\to D\pi$ and $B\to\pi\pi$.}

\section{Introduction}

Non-leptonic two-body decays of $B$ mesons, although simple as far as 
the underlying weak decay of the $b$ quark is concerned, are complicated 
on account of strong-interaction effects. If these effects could be 
computed, this would enhance tremendously our ability to uncover the 
origin of CP violation in weak interactions from data on a variety of 
such decays being collected at the $B$ factories. In these lecture, I 
review recent progress towards a systematic analysis of weak heavy-meson 
decays into two energetic mesons based on the factorization properties 
of decay amplitudes in QCD \cite{BBNS99,BBNS00}. My discussion will 
follow very closely the detailed account of this approach given in 
\cite{BBNS00}. (We have worked so hard on this paper that any attempt 
to improve on it were bound to fail and leave the author in despair.) 
Much of the credit for these notes belongs to my collaborators
Martin Beneke, Gerhard Buchalla, and Chris Sachrajda. 

As in the classic analysis of semi-leptonic $B\to D$ transitions 
\cite{IW89,VS87}, our arguments make extensive use of the fact that 
the $b$ quark is heavy compared to the intrinsic scale of strong 
interactions. This allows us to deduce that non-leptonic decay 
amplitudes in the heavy-quark limit have a simple structure. The 
arguments to reach this conclusion, however, are quite different from 
those used for semi-leptonic decays, since for non-leptonic decays a 
large momentum is transferred to at least one of the final-state 
mesons. The results of our work justify naive factorization of 
four fermion operators for many, but not all, non-leptonic decays and 
imply that corrections termed ``non-factorizable'', which up to now 
have been thought to be intractable, can be calculated rigorously if 
the mass of the decaying quark is large enough. This leads to 
a large number of predictions for CP-violating $B$ decays in the 
heavy-quark limit, for which measurements will soon become available.  

Weak decays of heavy mesons involve three fundamental scales, the 
weak-interaction scale $M_W$, the $b$-quark mass $m_b$, and the QCD 
scale $\Lambda_{\rm QCD}$, which are strongly ordered: 
$M_W\gg m_b\gg\Lambda_{\rm QCD}$. The underlying weak decay being 
computable, all theoretical work concerns strong-interaction 
corrections. QCD effects involving virtualities above the scale $m_b$ 
are well understood. They renormalize the coefficients of local 
operators $O_i$ in the effective weak Hamiltonian \cite{BBL}, so 
that the amplitude for the decay $B\to M_1 M_2$ is given by
\begin{equation}\label{effham}
   {\cal A}(B\to M_1 M_2) = \frac{G_F}{\sqrt2} \sum_i 
   \lambda_i\,C_i(\mu)\,\langle M_1 M_2 |O_i(\mu)|B\rangle \,,
\end{equation}
where each term in the sum is the product of a 
Cabibbo--Kobayashi--Maskawa (CKM) factor $\lambda_i$, a coefficient 
function $C_i(\mu)$, which incorporates strong-interaction effects 
above the scale $\mu\sim m_b$, and a matrix element of an operator 
$O_i$. The difficult theoretical problem is to compute these matrix 
elements or, at least, to reduce them to simpler non-perturbative 
objects. 

A variety of treatments of this problem exist, which rely on 
assumptions of some sort. Here we identify two somewhat contrary lines 
of approach. The first one, which we shall call ``naive 
factorization'', replaces the matrix element of a four-fermion operator 
in a heavy-quark decay by the product of the matrix elements of two 
currents \cite{FS78,CaMa78}, e.g.
\begin{equation}\label{fac1}
   \langle D^+\pi^-|(\bar c b)_{V-A}(\bar d  u)_{V-A}|\bar B_d\rangle
   \to \langle\pi^-|(\bar d u)_{V-A}|0\rangle\,
   \langle D^+|(\bar c b)_{V-A}|\bar B _d\rangle \,. 
\end{equation}
This assumes that the exchange of ``non-factorizable'' gluons between 
the $\pi^-$ and the $(\bar B_d\,D^+)$ system can be neglected if the 
virtuality of the gluons is below $\mu\sim m_b$. The non-leptonic decay 
amplitude then reduces to the product of a form factor and a decay 
constant. This assumption is in general not justified, except in the 
limit of a large number of colours in some cases. It deprives the 
amplitude of any physical mechanism that could account for rescattering 
in the final state. ``Non-factorizable'' radiative corrections must 
also exist, because the scale dependence of the two sides of 
(\ref{fac1}) is different. Since such corrections at scales larger than 
$\mu$ are taken into account in deriving the effective weak Hamiltonian, 
it appears rather arbitrary to leave them out below the scale $\mu$. 
Various generalizations of the naive factorization approach have been 
proposed, which include new parameters that account for 
non-factorizable corrections. In their most general form, these 
generalizations have nothing to do with the original ``factorization'' 
ansatz, but amount to a general parameterization of the matrix elements. 
Such general parameterizations are exact, but at the price of 
introducing many unknown parameters and eliminating any theoretical 
input on strong-interaction dynamics.

The second method used to study non-leptonic decays is the 
hard-scatter\-ing approach, which assumes the dominance of hard gluon 
exchange. The decay amplitude is then expressed as a convolution of a 
hard-scattering factor with light-cone wave functions of the 
participating mesons, in analogy with more familiar applications of 
this method to hard exclusive reactions involving only light hadrons 
\cite{LB80,EfRa80}. In many cases, the hard-scattering contribution 
represents the leading term in an expansion in powers of 
$\Lambda_{\rm QCD}/Q$, where $Q$ denotes the hard scale. However, the 
short-distance dominance of hard exclusive processes is not enforced 
kinematically and relies crucially on the properties of hadronic wave 
functions. There is an important difference between light mesons and 
heavy mesons in this regard, because the light quark in a heavy meson 
at rest naturally has a small momentum of order $\Lambda_{\rm QCD}$, 
while for fast light mesons a configuration with a soft quark is 
suppressed by the endpoint behaviour of the meson wave function. As a 
consequence, the soft (or Feynman) mechanism is power suppressed for 
hard exclusive processes involving light mesons, but it is of leading 
power for heavy-meson decays.

It is clear from this discussion that a satisfactory treatment should 
take into account soft contributions, but also allow us to compute 
corrections to naive factorization in a systematic way. It is not at 
all obvious that such a treatment would result in a predictive 
framework. We will show that this does indeed happen for most 
non-leptonic two-body $B$ decays. Our main conclusion is that 
``non-factorizable'' corrections are dominated by hard gluon exchange, 
while the soft effects that survive in the heavy-quark limit are 
confined to the $(B M_1)$ system, where $M_1$ denotes the meson that 
picks up the spectator quark in the $B$ meson. This result is expressed 
as a factorization formula, which is valid up to corrections suppressed 
by powers of $\Lambda_{\rm QCD}/m_b$. At leading power, 
non-perturbative contributions are parameterized by the physical form 
factors for the $B\to M_1$ transition and leading-twist light-cone 
distribution amplitudes of the mesons. Hard perturbative corrections 
can be computed systematically in a way similar to the hard-scattering 
approach. On the other hand, because the $B\to M_1$ transition is 
parameterized by a form factor, we recover the result of naive 
factorization at lowest order in $\alpha_s$. 

An important implication of the factorization formula is that strong 
rescattering phases are either perturbative or power suppressed in 
$\Lambda_{\rm QCD}/m_b$. It is worth emphasizing that the decoupling of 
$M_2$ occurs in 
the presence of soft interactions in the $(B M_1)$ system. In other 
words, while strong-interaction effects in the $B\to M_1$ transition 
are not confined to small transverse distances, the other meson $M_2$ 
is predominantly produced as a compact object with small transverse 
extension. The decoupling of soft effects then follows from ``colour 
transparency''. The colour-transparency argument for exclusive $B$ 
decays has already been noted in the literature \cite{Bj89,DG91}, but 
it has never been developed into a factorization formula that could be 
used to obtain quantitative predictions.

The approach described in \cite{BBNS99,BBNS00} is general and applies 
to decays into a heavy and a light meson (such as $B\to D\pi$) as well 
as to decays into two light mesons (such as $B\to\pi\pi$). 
Factorization does not hold, however, for decays such as 
$B\to\pi D$ and $B\to D\bar D$, in which the meson that does {\em not\/} 
pick up the spectator quark in the $B$ meson is heavy. For the main part 
in these lectures, we will focus on the case of $B\to D^{(*)} L$ decays 
(with $L$ a light meson), for which the factorization formula takes its 
simplest form, and power counting will be relatively straightforward.
Occasionally, we will point out what changes when we consider more
complicated decays such as $B\to\pi\pi$. A detailed treatment of these 
processes can be found in \cite{BBNSfuture}.

The outline of these notes is as follows: In Sect.~\ref{factform} we 
state the factorization formula in its general form. In 
Sect.~\ref{arguments} we collect the 
physical arguments that lead to factorization and introduce
our power-counting scheme. We show how light-cone distribution 
amplitudes enter, discuss the heavy-quark scaling of the $B\to D$ form 
factor, and explain the cancellation of soft and collinear contributions 
in ``non-factorizable'' vertex corrections to non-leptonic decay 
amplitudes. We also comment on the implications of our results for 
final-state interactions in hadronic $B$ decays. The cancellation of 
long-distance singularities is demonstrated in more detail in 
Sect.~\ref{oneloop}, where we present the calculation of the 
hard-scattering functions at one-loop order for decays into a heavy and 
a light meson. Various sources of power-suppressed effects, which give 
corrections to the factorization formula, are discussed in
Sect.~\ref{sec:power}. They include hard-scattering contributions, weak 
annihilation, and contributions from multi-particle Fock states. We 
then point out some limitations of the factorization approach. In 
Sect.~\ref{bdpi} we consider the phenomenology of $B\to D^{(*)} L$ 
decays on the basis of the factorization formula and discuss various 
tests of our theoretical framework. We also examine to what extent a 
charm meson should be considered as heavy or light. 
Section~\ref{conclusion} contains the conclusion.

\section{Statement of the factorization formula}
\label{factform}

In this section we summarize the factorization formula for non-leptonic 
$B$ decays. We introduce relevant terminology and 
definitions.

\subsection{The idea of factorization}
\label{idea}

In the context of non-leptonic decays the term ``factorization'' is 
usually applied to the approximation of the matrix element of a 
four-fermion operator by the product of a form factor and a decay 
constant, as illustrated in (\ref{fac1}). Corrections to this 
approximation are called ``non-factorizable''. We will refer to this 
approximation as ``naive factorization'' and use quotes on 
``non-factorizable'' to avoid confusion with the (much less trivial) 
meaning of factorization in the context of hard processes in QCD.
In the latter case, factorization refers to the separation of 
long-distance contributions to the process from a short-distance part 
that depends only on the large scale $m_b$. The short-distance part 
can be computed in an expansion in the strong coupling $\alpha_s(m_b)$. 
The long-distance contributions must be computed non-perturbatively or 
determined experimentally. The advantage is that these non-perturbative 
parameters are often simpler in structure than the original quantity, 
or they are process independent. For example, factorization applied to 
hard processes in inclusive hadron--hadron collisions requires only 
parton distributions as non-perturbative inputs. Parton distributions 
are much simpler objects than the original matrix element with two 
hadrons in the initial state. On the other hand, factorization applied 
to the $B\to D$ form factor leads to a non-perturbative object (the 
``Isgur--Wise function''), which is still a function of the momentum 
transfer. However, the benefit here is that symmetries relate this 
function to other form factors. In the case of non-leptonic $B$ decays, 
the simplification is primarily of the first kind (simpler structure). 
We call those effects non-factorizable (without quotes) which depend on 
the long-distance properties of the $B$ meson and both final-state 
mesons combined. 

The factorization properties of non-leptonic decay amplitudes depend on 
the two-meson final state. We call a meson ``light'' if its mass $m$ 
remains finite in the heavy-quark limit. A meson is called ``heavy'' if 
its mass scales with $m_b$ in the heavy-quark limit, 
such that $m/m_b$ stays fixed. In principle, we could still have 
$m\gg\Lambda_{\rm QCD}$ for a light meson. Charm mesons could be 
considered as light in this sense. However, unless otherwise mentioned, 
we assume that $m$ is of order $\Lambda_{\rm QCD}$ for a light meson, 
and we consider charm mesons as heavy. In evaluating the scaling 
behaviour of the decay amplitudes, we assume that the energies of both 
final-state mesons (in the $B$-meson rest frame) scale with $m_b$ in 
the heavy-quark limit. 

\subsection{The factorization formula}

\begin{figure}[t]
\vspace{-2.2cm}
\epsfxsize=14cm
\centerline{\epsffile{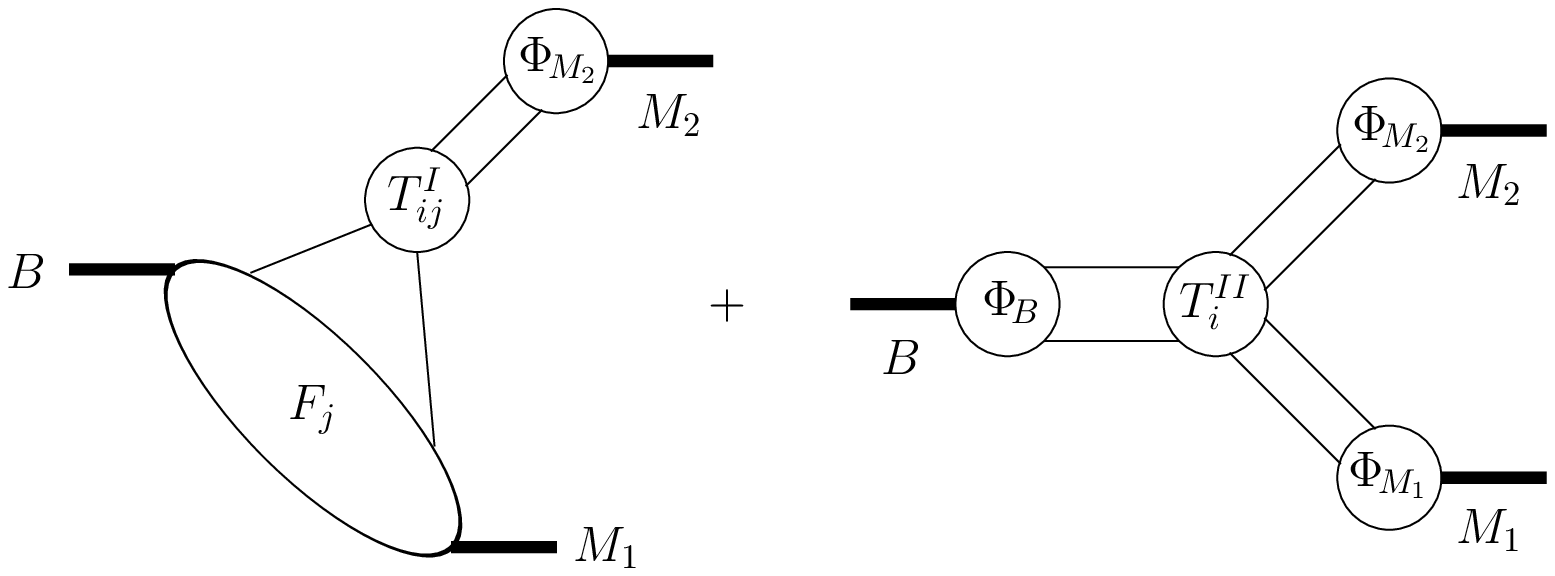}}
\vspace{-13.2cm}
\caption{\label{fig1}
Graphical representation of the factorization formula. Only one of the 
two form-factor terms in (\ref{fff}) is shown for simplicity.}
\end{figure}

We consider a generic weak decay $B\to M_1 M_2$ in the heavy-quark limit 
and differentiate between decays into final states containing a heavy 
and a light meson or two light mesons. Our goal is to show that, up to 
power corrections of order $\Lambda_{\rm QCD}/m_b$, the transition 
matrix element of an operator $O_i$ in the effective weak Hamiltonian 
can be written as
\begin{eqnarray}\label{fff}
   \langle M_1 M_2|O_i|B\rangle
   &=& \sum_j F_j^{B\to M_1}(m_2^2)\,f_{M_2} \int_0^1\!du\,
    T_{ij}^I(u)\,\Phi_{M_2}(u) \nonumber\\
   &&\mbox{if $M_1$ is heavy and $M_2$ is light,} \nonumber\\
   \langle M_1 M_2|O_i|B\rangle
   &=& \sum_j F_j^{B\to M_1}(m_2^2)\,f_{M_2} \int_0^1\!du\,
    T_{ij}^I(u)\,\Phi_{M_2}(u) ~+~ (M_1\leftrightarrow M_2) \nonumber\\
   &&\mbox{}+ f_B f_{M_1} f_{M_2} \int_0^1\!d\xi\,du\,dv\,
    T_i^{II}(\xi,u,v)\,\Phi_B(\xi)\,\Phi_{M_1}(v)\,\Phi_{M_2}(u)
    \nonumber\\
   &&\mbox{if $M_1$ and $M_2$ are both light.}
\end{eqnarray} 
Here $F_j^{B\to M}(m^2)$ denotes a $B\to M$ form factor evaluated at
$q^2=m^2$, $m_{1,2}$ are the light meson masses, and $\Phi_X(u)$ is 
the light-cone distribution amplitude for the quark--antiquark Fock 
state of the meson $X$. These non-perturbative quantities will be 
defined below. $T_{ij}^I(u)$ and $T_i^{II}(\xi,u,v)$ are hard-scattering 
functions, which are perturbatively calculable. The factorization 
formula in its general form is represented graphically in 
Fig.~\ref{fig1}.

The second equation in (\ref{fff}) applies to decays into two light 
mesons, for which the spectator quark in the $B$ meson (in the 
following simply referred to as the ``spectator quark'') can go to 
either of the final-state mesons. An example is the decay 
$B^-\to\pi^0 K^-$. If the spectator quark can go only to one of the 
final-state mesons, as for example in $\bar B_d\to\pi^+ K^-$, we call 
this meson $M_1$, and the second form-factor term on the right-hand 
side of (\ref{fff}) is absent. The formula simplifies when the 
spectator quark goes to a heavy meson (first equation in (\ref{fff})), 
such as in $\bar B_d\to D^+\pi^-$. Then the second term in 
Fig.~\ref{fig1}, which accounts for hard interactions with the 
spectator quark, can be dropped because it is power suppressed in the 
heavy-quark limit. In the opposite situation that the spectator quark 
goes to a light meson but the other meson is heavy, factorization does 
not hold, because the heavy meson is neither fast nor small and cannot 
be factorized from the $B\to M_1$ transition. Finally, notice that 
annihilation topologies do not appear in the factorization formula, 
since they do not contribute at leading order in the heavy-quark 
expansion. 

Any hard interaction costs a power of $\alpha_s$. As a consequence, the 
hard-spectator term in the second formula in (\ref{fff}) is absent at 
order $\alpha_s^0$. Since at this order the functions $T^I_{ij}(u)$ are 
independent of $u$, the convolution integral results in the 
normalization of the meson distribution amplitude, and (\ref{fff}) 
reproduces naive factorization. The factorization formula allows us to 
compute radiative corrections to this result to all orders in 
$\alpha_s$. Further corrections are suppressed by powers of 
$\Lambda_{\rm QCD}/m_b$ in the heavy-quark limit.

The significance and usefulness of the factorization formula stems 
from the fact that the non-perturbative quantities appearing on the 
right-hand side of the two equations in (\ref{fff}) are much simpler 
than the original non-leptonic matrix elements on the left-hand side. 
This is because they either reflect universal properties of a single 
meson (light-cone distribution amplitudes) or refer only to a 
$B\to\mbox{meson}$ transition matrix element of a local current (form 
factors). While it is extremely difficult, if not impossible 
\cite{MT90}, to compute the original matrix element 
$\langle M_1 M_2|O_i|B\rangle$ in lattice QCD, form factors and 
light-cone distribution amplitudes are already being computed in this 
way, although with significant systematic errors at present. 
Alternatively, form factors can be obtained using data on semi-leptonic 
decays, and light-cone distribution amplitudes by comparison with other 
hard exclusive processes.

After having presented the most general form of the factorization 
formula, we will from now on restrict ourselves to the case of 
heavy-light final states. Then the (simpler) first formula in 
(\ref{fff}) applies, and only the first term shown in Fig.~\ref{fig1} 
is present at leading power.

\subsection{Definition of non-perturbative parameters}

The form factors $F_j^{B\to M}(q^2)$ in (\ref{fff}) arise in the 
decomposition of current matrix elements of the form 
$\langle M(p')|\bar q\Gamma b|\bar B(p)\rangle$, where $\Gamma$ can be 
any irreducible Dirac matrix that appears after contraction of the hard 
subgraph to a local vertex with respect to the $B\to M$ transition. We 
will often refer to the matrix element of the vector current evaluated 
between a $B$ meson and a pseudoscalar meson $P$, which is 
conventionally parameterized as
\begin{eqnarray}
   \langle P(p')|\bar q\gamma^\mu b|\bar B(p)\rangle
   &=& F_+^{B\to P}(q^2)\,(p^\mu+{p'}^\mu) \nonumber\\
   &+& \Big[ F_0^{B\to P}(q^2) - F_+^{B\to P}(q^2) \Big]\,
    \frac{m_B^2-m_P^2}{q^2}\,q^\mu \,,
\end{eqnarray}
where $q=p-p'$, and $F_+^{B\to P}(0)=F_0^{B\to P}(0)$ at zero momentum
transfer. Note that we write (\ref{fff}) in terms of 
physical form factors. In principle, Fig.~\ref{fig1} could be looked 
upon in two different ways. We could suppose that the region represented 
by $F_j$ accounts only for the soft contributions to the $B\to M_1$ form 
factor. The hard contributions to the form  factor would then be 
considered as part of $T^{I}_{ij}$ (or as part of the second diagram). 
Performing this split-up would require that one understands the 
factorization of hard and soft contributions to the form factor. If 
$M_1$ is heavy, this amounts to matching the form factor onto a form 
factor defined in heavy-quark effective theory \cite{review}. However, 
for a light meson $M_1$ the factorization of hard and soft contributions 
to the form factor is not yet completely understood. We bypass this 
problem by interpreting $F_j$ as the physical form factor, including 
hard and soft contributions. This avoids the above problem, and in 
addition has the advantage that the physical form factors are directly 
related to measurable quantities.

Light-cone distribution amplitudes play the same role for hard 
exclusive processes that parton distributions play for inclusive 
processes. As in the latter case, the leading-twist distribution 
amplitudes, which are the ones we need at leading power in the $1/m_b$ 
expansion, are given by two-particle operators with a certain helicity 
structure. The helicity structure is determined by the angular 
momentum of the meson and the fact that the spinor of an energetic 
quark has only two large components. The leading-twist light-cone 
distribution amplitudes for pseudoscalar mesons ($P$) and 
longitudinally polarized vector mesons ($V_\parallel$) with flavour 
content $(\bar q q')$ are defined as
\begin{eqnarray}\label{distamps}
   \langle P(q)|\bar q(y)_\alpha q'(x)_\beta|0\rangle
   &=& \frac{i f_P}{4}\,(\not\!q\gamma_5)_{\beta\alpha}
    \int_0^1\!du\,e^{i(\bar{u} qx+u qy)}\,\Phi_P(u,\mu) \,, \nonumber\\
   \langle V_\parallel(q)|\bar q(y)_\alpha q'(x)_\beta|0\rangle
   &=& -\frac{i f_V}{4}\!\!\not\!q_{\beta\alpha} \int_0^1\!du\,
    e^{i(\bar{u} qx+u qy)}\,\Phi_\parallel(u,\mu) \,,
\end{eqnarray}
where $(x-y)^2=0$. We have suppressed the path-ordered exponentials that 
connect the two quark fields at different positions and make the 
light-cone operators gauge invariant. The equality sign is to be 
understood as ``equal up to higher-twist terms''. It is also understood 
that the operators on the left-hand side are colour singlets. When
convenient, we use the ``bar''-notation $\bar u\equiv 1-u$. The 
parameter $\mu$ is the renormalization scale of the light-cone operators 
on the left-hand side. The distribution amplitudes are normalized as 
$\int_0^1 du\,\Phi_X(u,\mu)=1$ with $X=P,V_\parallel$. One defines the 
asymptotic distribution amplitude as the limit in which the 
renormalization scale is sent to infinity. In this case
\begin{equation}\label{asform}
   \Phi_{X}(u,\mu) \stackrel{\mu\to\infty}{=} 6 u(1-u) \,.
\end{equation}

The use of light-cone distribution amplitudes in non-leptonic $B$ 
decays requires justification, which we will provide in 
Sects.~\ref{arguments} and \ref{oneloop}. The decay amplitude for a $B$ 
decay into a heavy-light final state is then calculated by assigning 
momenta $uq$ and $\bar u q$ to the quark and antiquark in the outgoing 
light meson (with momentum $q$), writing down the on-shell amplitude in 
momentum space, and performing the replacement
\begin{equation}\label{curproj}
   \bar u_{\alpha a}(u q)\,\Gamma(u,\dots)_{\alpha\beta,ab}
   v_{\beta b}(\bar u q) \to
   \frac{i f_P}{4 N_c} \int_0^1\!du\,\Phi_P(u)\,
   (\not\!q\gamma_5)_{\beta\alpha}\,\Gamma(u,\dots)_{\alpha\beta,aa}
\end{equation}
for pseudoscalars and, with obvious modifications, for vector mesons.
(Even when working with light-cone distribution amplitudes it is not 
always justified to perform the collinear approximation on the external 
quark and antiquark lines right away. One may have to keep the 
transverse components of the quark and antiquark momenta until after 
some operations on the amplitude have been carried out. However, these 
subtleties do not concern calculations at leading-twist order.)

\section{Arguments for factorization}
\label{arguments}

In this section we provide the basic power-counting arguments that 
lead to the factorized structure shown in (\ref{fff}). We do so by 
analyzing qualitatively the hard, soft and collinear contributions to 
the simplest Feynman diagrams.

\subsection{Preliminaries and power counting}

For concreteness, we label the charm meson which picks up the spectator 
quark by $M_1=D^+$ and assign momentum $p'$ to it. The light meson is 
labeled $M_2=\pi^-$ and assigned momentum $q=E\,n_+$, where $E$ is the
pion energy in the $B$ rest frame, and $n_\pm=(1,0,0,\pm 1)$ are 
four-vectors on the light-cone. At leading power, we neglect the mass 
of the light meson. 

The simplest diagrams that we can draw for a non-leptonic decay 
amplitude assign a quark and antiquark to each meson. We choose the 
quark and antiquark momenta in the pion as 
\begin{equation}\label{moms}
   l_q = u q + l_\perp + \frac{\vec l_\perp^{\,2}}{4uE}\,n_- \,,
   \qquad
   l_{\bar q} = \bar u q - l_\perp
   + \frac{\vec l_\perp^{\,2}}{4\bar uE}\,n_- \,.
\end{equation}
Note that $q\ne l_q+l_{\bar q}$, but the off-shellness 
$(l_q+l_{\bar q})^2$ is of the same order as the light meson mass, 
which we can neglect at leading power. A similar decomposition (with 
longitudinal momentum fraction $v$ and transverse momentum $l_\perp'$) 
is used for the charm meson.

To prove the factorization formula (\ref{fff}) for the case of 
heavy-light final states, one has to show that: 
\begin{itemize}
\item[i)] 
There is no leading (in powers of $\Lambda_{\rm QCD}/m_b$) contribution 
to the amplitude from the endpoint regions $u\sim\Lambda_{\rm QCD}/m_b$ 
and $\bar u\sim\Lambda_{\rm QCD}/m_b$.
\item[ii)]
One can set $l_\perp=0$ in the amplitude (more generally, expand the 
amplitude in powers of $l_\perp$) after collinear subtractions, which 
can be absorbed into the pion wave function. This, together with i), 
guarantees that the amplitude is legitimately expressed in terms of the 
light-cone distribution amplitudes of pion.
\item[iii)] 
The leading contribution comes from $\bar v\sim\Lambda_{\rm QCD}/m_b$ 
(the region where the spectator quark enters the charm meson as a soft 
parton), which guarantees the absence of a hard spectator interaction 
term. 
\item[iv)] 
After subtraction of infrared contributions corresponding to the 
light-cone distribution amplitude and the form factor, the leading 
contributions to the amplitude come only from internal lines with 
virtuality that scales with $m_b$.
\item[v)] 
Non-valence Fock states are non-leading.
\end{itemize}

The requirement that after subtractions virtualities should be large 
is obvious to guarantee the infrared finiteness of the hard-scattering 
functions $T^I_{ij}$. Let us comment on setting transverse momenta in 
the wave functions to zero and on endpoint contributions. Neglecting 
transverse momenta requires that we count them as order 
$\Lambda_{\rm QCD}$ when comparing terms of different magnitude in the 
scattering amplitude. This conforms to our intuition and the assumption 
of the parton model, that intrinsic transverse momenta are limited to 
hadronic scales. However, in QCD transverse momenta are not limited, 
but logarithmically distributed up to the hard scale. The important 
point is that contributions that violate the starting assumption 
of limited transverse momentum can be absorbed into the universal 
light-cone distribution amplitudes. The statement that transverse 
momenta can be counted of order $\Lambda_{\rm QCD}$ is to be understood 
after these subtractions have been performed.

The second comment concerns endpoint contributions in the convolution 
integrals over longitudinal momentum fractions. These contributions are 
dangerous, because we may be able to demonstrate the infrared safety of 
the hard-scattering amplitude under assumption of generic $u$ and 
independent of the shape of the meson distribution amplitude, but for 
$u\to 0$ or $u\to 1$ a propagator that was assumed to be off-shell 
approaches the mass-shell. If such a contribution were of leading power, 
we would not expect the perturbative calculation of the hard-scattering 
functions to be reliable.

Estimating endpoint contributions requires knowledge of the endpoint 
behaviour of the light-cone distribution amplitude. Since it 
enters the factorization formula at a renormalization scale of order 
$m_b$, we can use the asymptotic form (\ref{asform}) to estimate the 
endpoint contribution. (More generally, we only have to assume that the 
distribution amplitude at a given scale has the same endpoint behaviour 
as the asymptotic amplitude. This is generally the case, unless there 
is a conspiracy of terms in the Gegenbauer expansion of the 
distribution amplitude. If such a conspiracy existed at some scale, it 
would be destroyed by evolving the distribution amplitude to a 
different scale.) We count a light-meson distribution amplitude as order 
$\Lambda_{\rm QCD}/m_b$ in the endpoint region (defined as the region 
the quark or antiquark momentum is of order $\Lambda_{\rm QCD}$), and 
order $1$ away from the endpoint, i.e.\ (for $X=P,V_\parallel$)
\begin{equation}\label{powerpi}
   \Phi_{X}(u) \sim \left\{
   \begin{array}{cl}
    1 \,; & \quad \mbox{generic $u$,} \\[0.1cm]
    \Lambda_{\rm QCD}/m_b \,; & \quad
     u,\,\bar{u}\sim\Lambda_{\rm QCD}/m_b.
   \end{array}
   \right.
\end{equation}
Note that the endpoint region has a size of order 
$\Lambda_{\rm QCD}/m_b$, so that the endpoint suppression is 
$\sim(\Lambda_{\rm QCD}/m_b)^2$. This suppression has to be weighted 
against potential enhancements of the partonic amplitude when one of 
the propagators approaches the mass shell. The counting for $B$ mesons, 
or heavy mesons in general, is different. Naturally, the heavy quark 
carries almost all of the meson momentum, and hence we count
\begin{equation}\label{powerb}
   \Phi_B(\xi) \sim \left\{
   \begin{array}{cl}
    m_b/\Lambda_{\rm QCD} \,; & \quad \xi\sim\Lambda_{\rm QCD}/m_b,
    \\[0.1cm]
    0 \,; & \quad\xi\sim 1.
   \end{array}
   \right.
\end{equation}
The zero probability for a light spectator with momentum of order 
$m_b$ must be understood as a boundary condition for the wave function 
renormalized at a scale much below $m_b$. There is a small probability 
for hard fluctuations that transfer large momentum to the spectator. 
This ``hard tail'' is generated by evolution of the wave function from 
a hadronic scale to a scale of order $m_b$. If we assume that the 
initial distribution at the hadronic scale falls sufficiently rapidly 
for $\xi\gg \Lambda_{\rm QCD}/m_b$, this remains true after evolution. 
We shall assume a sufficiently fast fall-off, so that, for the purposes 
of power counting, the probability that the spectator-quark momentum 
is of order $m_b$ can be set to zero. The same counting applies to the 
$D$ meson. (Despite the fact that the charm meson has momentum of order 
$m_b$, we do not need to distinguish the rest frames of $B$ and $D$ for 
the purpose of power counting, because the two frames are not connected 
by a parametrically large boost. In other words, the components of the 
spectator quark in the $D$ meson are still of order 
$\Lambda_{\rm QCD}$.)

\subsection{The $B\to D$ form factor}
\label{formfactor}

We now demonstrate that the $B\to D$ form factor receives a leading 
contribution from soft gluon exchange. This implies that a non-leptonic 
decay cannot be treated completely in the hard-scattering picture, and 
so the form factor should enter the factorization formula as a 
non-perturbative quantity. 

\begin{figure}[t]
\vspace{-2.6cm}
\epsfxsize=18cm
\centerline{\epsffile{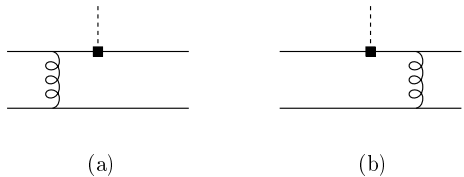}}
\vspace{-20.2cm}
\caption{\label{fig3}
Leading contributions to the $B\to D$ form factor in the 
hard-scattering approach. The dashed line represents the weak current. 
The two lines to the left belong to the $B$ meson, the ones to the 
right to the recoiling charm meson.}
\end{figure}

Consider the diagrams shown in Fig.~\ref{fig3}. When the exchanged 
gluon is hard the spectator quark in the final state has momentum of 
order $m_b$. But according to the counting rule (\ref{powerb}) this 
configuration has no overlap with the $D$-meson wave function. On the 
other hand, there is no suppression for soft gluons in Fig.~\ref{fig3}. 
It follows that the dominant behaviour of the $B\to D$ form factor in 
the heavy-quark limit is given by soft processes.

Because of this argument, we can exploit the heavy-quark symmetries
to determine how the form factor scales in the heavy-quark limit. The
well-known result is that the form factor scales like a constant (modulo
logarithms), since it is equal to one at zero velocity transfer and 
independent of $m_b$ as long as the Lorentz boost that connects the 
$B$ and $D$ rest frames is of order 1. The same conclusion follows from 
the power-counting rules for light-cone wave functions. To see this, we 
represent the form factor by an overlap integral of wave functions (not 
integrated over transverse momentum),
\begin{equation}\label{overlap1}
   F_{+,0}^{B\to D}(0) \sim 
   \int\frac{d\xi d^2k_\perp}{16\pi^3}\,\Psi_B(\xi,k_\perp)\,
   \Psi_D(\xi'(\xi),k_\perp) \,,
\end{equation}
where $\xi'(\xi)$ is fixed by kinematics, and we have set $q^2=0$ for
simplicity. The probability of finding the $B$ meson in its valence Fock 
state is of order 1 in the heavy-quark limit, i.e.\ 
\begin{equation}\label{normkt}
   \int\frac{d\xi d^2k_\perp}{16\pi^3}\,|\Psi_{B,D}(\xi,k_\perp)|^2
   \sim 1 \,.
\end{equation}
Counting $k_\perp\sim\Lambda_{\rm QCD}$ and 
$d\xi\sim\Lambda_{\rm QCD}/m_b$, we deduce that 
$\Psi_B(\xi,k_\perp)\sim m_b^{1/2}/\Lambda_{\rm QCD}^{3/2}$. From 
(\ref{overlap1}), we then obtain the scaling law 
$F_{+,0}^{B\to D}(0)\sim 1$, in agreement with the prediction of 
heavy-quark symmetry.

The representation (\ref{overlap1}) of the form factor as an overlap 
of wave functions for the two-particle Fock state of the heavy meson is 
not rigorous, because there is no reason to assume that the contribution 
from higher Fock states with additional soft gluons is suppressed. The 
consistency with the estimate based on heavy-quark symmetry shows that 
these additional contributions are not larger than the two-particle 
contribution.

\subsection{Non-leptonic decay amplitudes}
\label{nlamp}

We now turn to a qualitative discussion of the lowest-order and 
one-gluon exchange diagrams that could contribute to the 
hard-scattering kernels $T^I_{ij}(u)$ in (\ref{fff}). In the figures 
which follow, the two lines directed upwards represent $\pi^-$, the 
lines on the left represent $\bar B_d$, and the lines on the right 
represent $D^+$.

\subsubsection{Lowest-order diagram}

\begin{figure}[t]
\vspace{-4.1cm}
\epsfxsize=18cm
\centerline{\epsffile{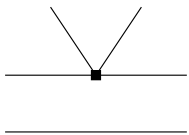}}
\vspace*{-19.5cm}
\caption{\label{fig4}
Leading-order contribution to the hard-scattering kernels $T^I_{ij}(u)$. 
The weak decay of the $b$ quark through a four-fermion operator is 
represented by the black square.}
\end{figure}

There is a single diagram with no hard gluon interactions shown in 
Fig.~\ref{fig4}. According to (\ref{powerb}) the spectator quark is 
soft, and since it does not undergo a hard interaction it is absorbed 
as a soft quark by the recoiling meson. This is evidently a 
contribution to the left-hand diagram of Fig.~\ref{fig1}, involving 
the $B\to D$ form factor. The hard subprocess in Fig.~\ref{fig4} is 
just given by the insertion of a four-fermion operator, and hence it   
does not depend on the longitudinal momentum fraction $u$ of the two 
quarks that form the emitted $\pi^-$. Consequently, the lowest-order 
contribution to $T_{ij}^I(u)$ in (\ref{fff}) is independent of $u$, 
and the $u$-integral reduces to the normalization condition for the 
pion distribution amplitude. The result is, not surprisingly, that the 
factorization formula reproduces the result of naive 
factorization if we neglect gluon exchange. Note that the physical 
picture underlying this lowest-order process is that the spectator 
quark (which is part of the $B\to D$ form factor) is soft. If this is 
the case, the hard-scattering approach misses the leading contribution 
to the non-leptonic decay amplitude. 

Putting together all factors relevant to power counting, we find that 
in the heavy-quark limit the decay amplitude for a decay into a 
heavy-light final state (in which the spectator quark is absorbed by 
the heavy meson) scales as 
\begin{equation}\label{abd}
   {\cal A}(\bar B_d\to D^+\pi^-)\sim G_F m_b^2\,F^{B\to D}(0)\,f_\pi 
   \sim G_F m_b^2\,\Lambda_{\rm QCD} \,.
\end{equation}
Other contributions must be compared with this scaling rule.

\subsubsection{Factorizable diagrams}

\begin{figure}[t]
\vspace{-2.8cm}
\epsfxsize=15cm
\centerline{\epsffile{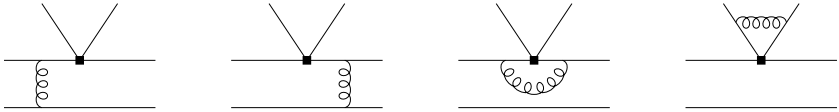}}
\vspace{-16.7cm}
\caption{\label{fig5} 
Diagrams at order $\alpha_s$ that need not be calculated.}
\end{figure}

In order to justify naive factorization as the leading term in an 
expansion in $\alpha_s$ and $\Lambda_{\rm QCD}/m_b$, we must show that 
radiative corrections are either suppressed in one of these two 
parameters, or already contained in the definition of the form factor 
and the pion decay constant. Consider the graphs shown in 
Fig.~\ref{fig5}. The first three diagrams are part of the form factor 
and do not contribute to the hard-scattering kernels. Since the first 
and third diagrams contain leading contributions from the region in 
which the gluon is soft, they should not be considered as corrections 
to Fig.~\ref{fig4}. However, this is of no consequence since these soft 
contributions are absorbed into the physical form factor. 

The fourth diagram in Fig.~\ref{fig5} is also factorizable. In general, 
this graph would split into a hard contribution and a contribution to 
the evolution of the pion distribution amplitude. However, as the 
leading-order diagram in Fig.~\ref{fig4} involves only the normalization 
integral of the pion distribution amplitude, the sum of the fourth 
diagram in Fig.~\ref{fig5} and the wave-function renormalization of the 
quarks in the emitted pion vanishes. In other words, these diagrams 
would renormalize the $(\bar u d)$ light-quark current, which however 
is conserved.

\subsubsection{``Non-factorizable'' vertex corrections}

We now begin the analysis of  ``non-factorizable'' diagrams, i.e.\ 
diagrams containing gluon exchanges that cannot be associated with the 
$B\to D$ form factor or the pion decay constant. At order $\alpha_s$, 
these diagrams can be divided into three groups: vertex corrections, 
hard spectator interactions, and annihilation diagrams. 

\begin{figure}[t]
\vspace{-3.3cm}
\epsfxsize=15cm
\centerline{\epsffile{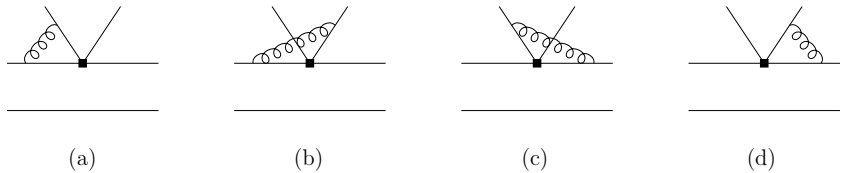}}
\vspace{-15.6cm}
\caption{\label{fig6}
``Non-factorizable'' vertex corrections.}
\end{figure}

The vertex corrections shown in Fig.~\ref{fig6} violate the naive 
factorization ansatz (\ref{fac1}). One of the key observations made in
\cite{BBNS99,BBNS00} is that these diagrams are calculable nonetheless. 
Let us summarize the argument here, postponing the explicit evaluation 
of these diagrams to Sect.~\ref{oneloop}. The statement is that the 
vertex-correction diagrams form an order-$\alpha_s$ contribution to the 
hard-scattering kernels $T^I_{ij}(u)$. To demonstrate this, we have to 
show that: i) The transverse momentum of the quarks that form the pion
can be neglected at leading power, i.e.\ the two momenta in (\ref{moms}) 
can be approximated by $u q$ and $\bar u q$, respectively. This 
guarantees that only a convolution in the longitudinal momentum fraction 
$u$ appears in the factorization formula. ii) The contribution from the 
soft-gluon region and gluons collinear to the direction of the pion is 
power suppressed. In practice, this means that the sum of these diagrams 
cannot contain any infrared divergences at leading power in 
$\Lambda_{\rm QCD}/m_b$.

Neither of the two conditions holds true for any of the four diagrams 
individually, as each of them separately contains collinear and infrared 
divergences. As will be shown in detail later, the infrared divergences 
cancel when one sums over the gluon attachments to the two quarks 
comprising the emission pion ((a+b), (c+d) in Fig.~\ref{fig6}). This 
cancellation is a technical manifestation of Bjorken's 
colour-transparency argument \cite{Bj89}, stating that soft gluon 
interactions with the emitted colour-singlet $(\bar u d)$ pair are 
suppressed because they interact with the colour dipole moment of the 
compact light-quark pair. Collinear divergences cancel after summing 
over gluon attachments to the $b$ and $c$ quark lines ((a+c), (b+d) in 
Fig.~\ref{fig6}). Thus the sum of the four diagrams (a--d) involves 
only hard gluon exchange at leading power. Because the hard gluons 
transfer large momentum to the quarks that form the emission pion, the 
hard-scattering factor now results in a non-trivial convolution with 
the pion distribution amplitude. ``Non-factorizable'' contributions are 
therefore non-universal, i.e.\ they depend on the quantum numbers of 
the final-state mesons.

Note that the colour-transparency argument, and hence the cancellation 
of soft gluon effects, applies only if the $(\bar u d)$ pair is compact. 
This is not the case if the emitted pion is formed in a very 
asymmetric configuration, in which one of the quarks carries almost all 
of the pion momentum. Since the probability for forming a pion in 
such an endpoint configuration is of order $(\Lambda_{\rm QCD}/m_b)^2$, 
they could become important only if the hard-scattering amplitude 
favoured the production of these asymmetric pairs, i.e.\ if 
$T^I_{ij}\sim 1/u^2$ for $u\to 0$ (or $T^I_{ij}\sim 1/\bar u^2$ for 
$u\to 1$). However, we will see that such strong endpoint singularities 
in the hard-scattering amplitude do not occur.

To complete the argument, we have to show that all other types of 
contributions to the non-leptonic decay amplitudes are power suppressed
in the heavy-quark limit. This includes interactions with the spectator
quark, weak annihilation graphs, and contributions from higher Fock 
components of the meson wave functions. This will be done in 
Sect.~\ref{sec:power}. In summary, then, for hadronic $B$ decays into a 
light emitted and a heavy recoiling meson the first factorization 
formula in (\ref{fff}) holds. At order $\alpha_s$, the hard-scattering 
kernels $T^I_{ij}(u)$ are computed from the diagrams shown in 
Figs.~\ref{fig4} and \ref{fig6}. Naive factorization follows when one 
neglects all corrections of order $\Lambda_{\rm QCD}/m_b$ {\em and\/} 
$\alpha_s$. The factorization formula allows us to compute 
systematically corrections to higher order in $\alpha_s$, but still 
neglects power corrections.

\subsection{Remarks on final-state interactions}
\label{fsi}

Some of the loop diagrams entering the calculation of the 
hard-scattering kernels have imaginary parts, which contribute to the 
strong rescattering phases. It follows from our discussion that these 
imaginary parts are of order $\alpha_s$ or $\Lambda_{\rm QCD}/m_b$. 
This demonstrates that strong phases vanish in the heavy-quark limit 
(unless the real parts of the amplitudes are also suppressed). Since 
this statement goes against the folklore that prevails from the present 
understanding of this issue, and since the subject of final-state 
interactions (and of strong-interaction phases in particular) is of 
paramount importance for the interpretation of CP-violating 
observables, a few additional remarks are in order.

Final-state interactions are usually discussed in terms of intermediate 
hadronic states. This is suggested by the unitarity relation (taking 
$B\to \pi\pi$ for definiteness) 
\begin{equation}\label{unitarity}
   \mbox{Im}\,{\cal A}_{B\to \pi\pi}\sim \sum_n 
   {\cal A}_{B\to n}\,{\cal A}_{n\to\pi\pi}^* \,,
\end{equation}
where $n$ runs over all hadronic intermediate states. We can also 
interpret the sum in (\ref{unitarity}) as extending over intermediate 
states of partons. The partonic interpretation is justified by the 
dominance of hard rescattering in the heavy-quark limit. In this limit, 
the number of physical intermediate states is arbitrarily large. We 
may then argue on the grounds of parton--hadron duality that their 
average is described well enough (up to $\Lambda_{\rm QCD}/m_b$ 
corrections, say) by a partonic calculation. This is the picture 
implied by (\ref{fff}). The hadronic language is in principle exact. 
However, the large number of intermediate states makes it intractable 
to observe systematic cancellations, which usually occur in an 
inclusive sum over hadronic intermediate states.

A particular contribution to the right-hand side of (\ref{unitarity}) 
is elastic rescattering ($n=\pi\pi$). The energy dependence of the 
total elastic $\pi\pi$-scattering cross section is governed by soft 
pomeron behaviour. Hence the strong-interaction phase of the 
$B\to\pi\pi$ amplitude due to elastic rescattering alone increases 
slowly in the heavy-quark limit \cite{DGPS96}. On general grounds, it 
is rather improbable that elastic rescattering gives an appropriate 
representation of the imaginary part of the decay amplitude in the 
heavy-quark limit. This expectation is also borne out in the framework 
of Regge behaviour, as discussed in \cite{DGPS96}, where the importance 
(in fact, dominance) of inelastic rescattering was emphasized. However, 
this discussion left open the possibility of soft rescattering phases 
that do not vanish in the heavy-quark limit, as well as the possibility 
of systematic cancellations, for which the Regge approach does not 
provide an appropriate theoretical framework.

Eq.~(\ref{fff}) implies that such systematic cancellations {\em do\/} 
occur in the sum over all intermediate states $n$. It is worth recalling 
that similar cancellations are not uncommon for hard processes. Consider 
the example of $e^+ e^-\to\,$hadrons at large energy $q$. While the 
production of any hadronic final state occurs on a time scale of order 
$1/\Lambda_{\rm QCD}$ (and would lead to infrared divergences if we 
attempted to describe it using perturbation theory), the inclusive 
cross section given by the sum over all hadronic final states is 
described very well by a $(q\bar q)$ pair that lives over a short time 
scale of order $1/q$. In close analogy, while each particular hadronic 
intermediate state $n$ in (\ref{unitarity}) cannot be described 
partonically, the sum over all intermediate states is accurately 
represented by a $(q\bar q)$ fluctuation of small transverse size of 
order $1/m_b$. Because the $(q\bar q)$ pair is small, the physical 
picture of rescattering is very different from elastic $\pi\pi$ 
scattering.

In perturbation theory, the pomeron is associated with two-gluon 
exchange. The analysis of two-loop contributions to the non-leptonic 
decay amplitude in \cite{BBNS00} shows that the soft and collinear 
cancellations that guarantee the partonic interpretation of 
rescattering extend to two-gluon exchange. Hence, the soft final-state 
interactions are again subleading as required by the validity of 
(\ref{fff}). As far as the hard rescattering contributions are 
concerned, two-gluon exchange plus ladder graphs between a compact 
$(q\bar q)$ pair with energy of order $m_b$ and transverse size of order 
$1/m_b$ and the other pion does not lead to large logarithms, and 
hence there is no possibility to construct the (hard) pomeron. Note 
the difference with elastic vector-meson production through a virtual 
photon, which also involves a compact $(q\bar q)$ pair. However, in 
this case one considers $s\gg Q^2$, where $\sqrt{s}$ is the 
photon--proton center-of-mass energy and $Q$ the virtuality of the 
photon. This implies that the $(q\bar q)$ fluctuation is born long 
before it hits the proton. It is this difference of time scales, 
non-existent in non-leptonic $B$ decays, that permits pomeron exchange 
in elastic vector-meson production in $\gamma^* p$ collisions.

\boldmath
\section{$B\to D\pi$: Factorization at one-loop order}
\unboldmath
\label{oneloop}

We now present a more detailed treatment of the exclusive decays 
$\bar B_d\to D^{(*)+} L^-$, where $L$ is a light meson. We illustrate 
explicitly how factorization emerges at one-loop order and compute the 
hard-scattering kernels $T_{ij}^I(u)$ in the factorization formula 
(\ref{fff}). For each final state $f$, we express the decay amplitudes 
in terms of parameters $a_{1}(f)$ defined in analogy with similar 
parameters used in the literature on naive factorization.

\subsection{Effective Hamiltonian and decay topologies}

The effective Hamiltonian for $B\to D\pi$ is
\begin{equation}\label{heff18}
   {\cal H}_{\rm eff} = \frac{G_F}{\sqrt{2}}\,V^*_{ud}V_{cb}\,
   ( C_0 O_0 + C_8 O_8 ) \,.
\end{equation}
We choose to write the two independent four-quark operators in the 
singlet--octet basis
\begin{eqnarray}\label{o18}
   O_0 &=& \bar c\gamma^\mu(1-\gamma_5)b\, 
    \bar d\gamma_\mu(1-\gamma_5)u \,, \nonumber\\
   O_8 &=& \bar c\gamma^\mu(1-\gamma_5)T^A b\, 
    \bar d\gamma_\mu(1-\gamma_5)T^A u \,,
\end{eqnarray}
rather than in the more conventional basis of $O_1$ and $O_2$. The 
Wilson coefficients $C_0$ and $C_8$ describe the exchange of hard 
gluons with virtualities between the high-energy matching scale $M_W$ 
and a renormalization scale $\mu$ of order $m_b$. (These coefficients 
are related to the ones of the standard basis by $C_0=C_1+C_2/3$ and 
$C_8=2C_2$.) They are known at next-to-leading order in 
renormalization-group improved perturbation theory and are given by 
\cite{BBL}
\begin{equation}\label{c18}
   C_0 = \frac{N_c+1}{2N_c}\,C_+ + \frac{N_c-1}{2N_c}\,C_- \,,\qquad
   C_8 = C_+ - C_- \,,
\end{equation}
where
\begin{equation}\label{cpm}
   C_\pm(\mu) = \left( 1 + \frac{\alpha_s(\mu)}{4\pi}\,B_\pm \right)\,
   \bar C_\pm(\mu) \,, \qquad
   B_\pm = \pm\frac{N_c\mp 1}{2N_c}\,B \,,
\end{equation}
and
\begin{equation}\label{cpmb}
   \bar C_\pm(\mu) = \left[ \frac{\alpha_s(M_W)}{\alpha_s(\mu)}
   \right]^{d_\pm} \left[ 1
   + \frac{\alpha_s(M_W)-\alpha_s(\mu)}{4\pi}\,S_\pm \right] \,.
\end{equation}
For $N_c=3$ and $f=5$, we have $d_+=\frac{6}{25}$ and 
$d_-=-\frac{12}{25}$, as well as $S_+=\frac{6473}{3174}$ and 
$S_-=-\frac{9371}{1587}$. The scheme dependence of the Wilson 
coefficients at next-to-leading order is parameterized by the 
coefficient $B$ in (\ref{cpm}). We note that $B_{\rm NDR}=11$ in the 
naive dimensional regularization (NDR) scheme with anticommuting 
$\gamma_5$, and  $B_{\rm HV}=7$ in the `t~Hooft--Veltman (HV) scheme. 
We will demonstrate below that the scale and scheme dependence of the 
Wilson coefficients is canceled by a corresponding scale and scheme 
dependence of the hadronic matrix elements of the operators $O_0$ and 
$O_8$.

\begin{figure}[t]
\epsfxsize=8cm
\centerline{\epsffile{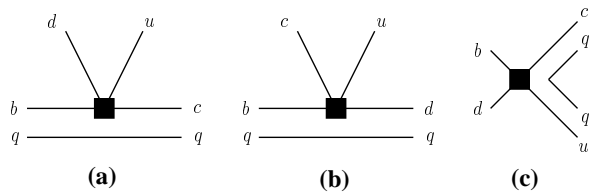}}
\caption{\label{fig:bdpi}
Basic quark-level topologies for $B\to D\pi$ decays ($q=u$, $d$): (a) 
class-I, (b) class-II, (c) weak annihilation. $\bar B_d\to D^+\pi^-$ 
receives contributions from (a) and (c), $\bar B_d\to D^0\pi^0$ from
(b) and (c), and $B^-\to D^0\pi^-$ from (a) and (b). Only (a) 
contributes in the heavy-quark limit.}
\end{figure}

Before continuing with a discussion of these matrix elements, it is 
useful to consider the flavour structure for the various contributions 
to $B\to D\pi$ decays. The possible quark-level topologies are depicted 
in Fig.~\ref{fig:bdpi}. In the terminology generally adopted for
two-body non-leptonic decays, the decays $\bar B_d\to D^+\pi^-$,
$\bar B_d\to D^0\pi^0$ and $B^-\to D^0\pi^-$ are referred to as 
class-I, class-II and class-III, respectively \cite{NeSt97}. In 
$\bar B_d\to D^+\pi^-$ and $B^-\to D^0\pi^-$ decays the pion can be 
directly created from the weak current. We call this a class-I 
contribution, following the above terminology. In addition, in the case 
of $\bar B_d\to D^+\pi^-$ there is a contribution from weak 
annihilation, and a class-II amplitude contributes to $B^-\to D^0\pi^-$. 
The important point is that the spectator quark goes into the light 
meson in the case of the class-II amplitude. This amplitude is 
suppressed in the heavy-quark limit, as is the annihilation amplitude. 
It follows that the amplitude for $\bar B_d\to D^0\pi^0$, receiving only 
class-II and annihilation contributions, is subleading compared with 
$\bar B_d\to D^+\pi^-$ and $B^-\to D^0\pi^-$, which are dominated by 
the class-I topology.

We shall use the one-loop analysis for $\bar B_d\to D^+\pi^-$ as a
concrete example to illustrate explicitly the various steps involved in 
establishing the factorization formula. Most of the arguments given 
below are standard from the theory of hard exclusive processes 
involving light hadrons \cite{LB80}. However, it is instructive to 
repeat these arguments in the context of $B$ decays.

\subsection{Soft and collinear cancellations at one-loop order}
\label{oneloopcancel}

In order to demonstrate the property of factorization for the decay
$\bar B_d\to D^+\pi^-$, we now analyze the ``non-factorizable''
one-gluon exchange contributions shown in Fig.~\ref{fig6} in some 
detail. We consider the leading, valence Fock state of the emitted 
pion. This is justified since higher Fock components only give 
power-suppressed contributions to the decay amplitude in the 
heavy-quark limit (as demonstrated later). For the purpose of our
discussion, the valence Fock state of the pion can be written as
\begin{equation}\label{piwf}
   |\pi(q)\rangle = \int\frac{du}{\sqrt{u\bar u}}\,
   \frac{d^2 l_\perp}{16\pi^3}\,\frac{1}{\sqrt{2 N_c}}
   \left( a^\dagger_\uparrow(l_q)\,b^\dagger_\downarrow(l_{\bar q})
   - a^\dagger_\downarrow(l_q)\,b^\dagger_\uparrow(l_{\bar q}) \right)
   |0\rangle\,\Psi(u,\vec l_\perp) \,,
\end{equation}
where $a^\dagger_s$ ($b^\dagger_s$) denotes the creation operator for a 
quark (antiquark) in a state with spin $s=\uparrow$ or $s=\downarrow$, 
and we have suppressed colour indices. The wave function 
$\Psi(u,\vec l_\perp)$ is defined as the amplitude for the pion to be 
composed of two on-shell quarks, characterized by longitudinal momentum 
fraction $u$ and transverse momentum $l_\perp$. The on-shell momenta
of the quark and antiquark are chosen as in (\ref{moms}). For 
the purpose of power counting, $l_\perp\sim\Lambda_{\rm QCD}\ll E
\sim m_b$. Note that the invariant mass of the valence state is 
$(l_q+l_{\bar q})^2=\vec l^{\,2}_\perp/(u\bar u)$, which is of order
$\Lambda^2_{\rm QCD}$ and hence negligible in the heavy-quark limit
unless $u$ is in the vicinity of the endpoints $u=0$ or 1. In this 
case, the invariant mass of the quark--antiquark pair becomes large, 
and the valence Fock state is no longer a valid representation of the 
pion. However, in the heavy-quark limit the dominant contributions to 
the decay amplitude come from configurations where both partons are 
hard ($u$ and $\bar u$ both of order 1), and so the two-particle Fock 
state yields a consistent description. We will provide an explicit 
consistency check of this important feature later on.

As a next step, we write down the amplitude
\begin{equation}\label{piudb}
   \langle\pi(q)|u(0)_\alpha\bar d(y)_\beta|0\rangle
   = \int\!du\,\frac{d^2 l_\perp}{16\pi^3}\,\frac{1}{\sqrt{2 N_c}}\,
   \Psi^*(u,\vec l_\perp)\,(\gamma_5\!\not\!q)_{\alpha\beta}\,
   e^{i l_q\cdot y} \,,
\end{equation}
which appears as an ingredient of the $B\to D\pi$ matrix element. 
It is now straightforward to obtain the one-gluon exchange 
contribution to the $B\to D\pi$ matrix element of the operator $O_8$. 
For the sum of the four diagrams in Fig.~\ref{fig6}, we find
\begin{eqnarray}\label{o8a1a2}
   \langle D^+\pi^-|O_8|\bar B_d\rangle_{\rm 1\mbox{-}gluon} &=& \\
   &&\hspace*{-3.7cm}
    i g_s^2\frac{C_F}{2}\!\int\!\frac{d^4k}{(2\pi)^4}\,
    \langle D^+|\bar c A_1(k) b|\bar B_d\rangle\,\frac{1}{k^2}\!
    \int^1_0\!du\,\frac{d^2 l_\perp}{16\pi^3}\,
    \frac{\Psi^*(u,\vec l_\perp)}{\sqrt{2 N_c}}\,
    {\rm tr}[\gamma_5\!\not\!q A_2(l_q,l_{\bar q},k)] \,, \nonumber
\end{eqnarray}
where
\begin{eqnarray}\label{a1bc}
   A_1(k)
   &=& \frac{\gamma^\lambda(\not\!p_c-\not\!k+m_c)\Gamma}
            {2p_c\cdot k-k^2}
    - \frac{\Gamma(\not\!p_b+\not\!k+m_b)\gamma^\lambda}
           {2p_b\cdot k+k^2} \,, \nonumber\\
   A_2(l_q,l_{\bar{q}},k)
   &=& \frac{\Gamma(\not\!l_{\bar q}+\not\!k)\gamma_\lambda}
            {2l_{\bar q}\cdot k+k^2}
    - \frac{\gamma_\lambda(\not\!l_q+\not\!k)\Gamma}
           {2l_q\cdot k+k^2} \,.
\end{eqnarray}
Here $\Gamma=\gamma^\mu(1-\gamma_5)$, and $p_b$, $p_c$ are the momenta 
of the $b$- and $c$-quark, respectively. There is no correction 
to the matrix element of $O_0$ at order $\alpha_s$, because in this case 
the $(d\bar u)$ pair is necessarily in a colour-octet configuration and 
cannot form a pion.

In (\ref{o8a1a2}) the pion wave function $\Psi(u,l_\perp)$ appears 
separated from the $B\to D$ transition. This is merely a reflection of 
the fact that we have represented the pion state in the form shown in 
(\ref{piwf}). It does not, by itself, imply factorization, since the 
right-hand side of (\ref{o8a1a2}) still involves non-trivial 
integrations over $\vec l_\perp$ and the gluon momentum $k$, and long- 
and short-distance contributions are not yet disentangled. In order to
prove factorization, we need to show that the integral over $k$ 
receives only subdominant contributions from the region of small $k^2$. 
This is equivalent to showing that the integral over $k$ does not 
contain infrared divergences at leading power in $\Lambda_{\rm QCD}/m_b$. 

To demonstrate infrared finiteness of the one-loop integral
\begin{equation}\label{dka1a2}
   J \equiv \int\!d^4k\,\frac{1}{k^2}\,
   A_1(k)\otimes A_2(l_q,l_{\bar{q}},k)
\end{equation}
at leading power, the heavy-quark limit and the corresponding large 
light-cone momentum of the pion are again essential. First note that 
when $k$ is of order $m_b$, $J\sim 1$ by dimensional analysis. 
Potential infrared divergences could arise when $k$ is soft or 
collinear to the pion momentum $q$. We need to show that the 
contributions from these regions are power suppressed. (Note 
that we do not need to show that $J$ is infrared finite. It is enough 
that logarithmic divergences have coefficients that are power 
suppressed.)

We treat the soft region first. Here all components of $k$ become small 
simultaneously, which we describe by scaling $k\sim\lambda$. Counting 
powers of $\lambda$ ($d^4k\sim \lambda^4$, $1/k^2\sim\lambda^{-2}$,
$1/p\cdot k\sim\lambda^{-1}$) reveals that each of the four diagrams
in Fig.~\ref{fig6}, corresponding to the four terms in the product in 
(\ref{dka1a2}), is logarithmically divergent. However, because $k$ is 
small the integrand can be simplified. For instance, the second term 
in $A_2$ can be approximated as
\begin{equation}\label{a2lperp}
   \frac{\gamma_\lambda(\not\!l_q+\not\!k)\Gamma}{2l_q\cdot k+k^2}
   = \frac{\gamma_\lambda(\!u\not\!q+\not\!l_\perp
           + \frac{\vec l^2_\perp}{4uE} \not\!n_- +\not\!k)\Gamma}
          {2u q\cdot k+2l_\perp\cdot k+\frac{\vec l^2_\perp}{2uE}
           n_-\cdot k+k^2}
   \simeq \frac{q_\lambda}{q\cdot k}\,\Gamma \,,
\end{equation}
where we used that $\!\not\!q$ to the extreme left or right of an 
expression gives zero due to the on-shell condition for the external 
quark lines. We get exactly the same expression but with an opposite 
sign from the other term in $A_2$, and hence the soft divergence cancels 
out. More precisely, we find that the integral is infrared finite in 
the soft region when $l_\perp$ is neglected. When $l_\perp$ is not 
neglected, there is a divergence from soft $k$ which is proportional to 
$l^2_\perp/m_b^2\sim\Lambda^2_{\rm QCD}/m_b^2$. In either case, the 
soft contribution to $J$ is of order $\Lambda_{\rm QCD}/m_b$ or smaller 
and hence suppressed relative to the hard contribution. This corresponds 
to the standard soft cancellation mechanism, which is a technical 
manifestation of colour transparency.

Each of the four terms in (\ref{dka1a2}) is also divergent when $k$ 
becomes collinear with the light-cone momentum $q$. This implies the 
scaling $k^+\sim\lambda^0$, $k_\perp\sim\lambda$, and $k^-\sim\lambda^2$.
Then $d^4k\sim dk^+ dk^- d^2k_\perp\sim\lambda^4$, and
$q\cdot k=q^+ k^-\sim\lambda^2$, $k^2=2k^+ k^-+k^2_\perp\sim\lambda^2$.
The divergence is again logarithmic, and it is thus sufficient to 
consider the leading behaviour in the collinear limit. Writing 
$k=\alpha q+\ldots$ we can now simplify the second term of $A_2$ as
\begin{equation}\label{a2lperp2}
   \frac{\gamma_\lambda(\not\!l_q+\not\!k)\Gamma}{2 l_q\cdot k+k^2}
   \simeq q_\lambda\,\frac{2(u+\alpha)\Gamma}{2l_q\cdot k+k^2} \,.
\end{equation}
No simplification occurs in the denominator (in particular, $l_\perp$ 
cannot be neglected), but the important point is that the leading 
contribution is proportional to $q_\lambda$. Therefore, substituting 
$k=\alpha q$ into $A_1$ and using $q^2=0$, we obtain 
\begin{equation}\label{a1coll}
   q_\lambda A_1\simeq
   \frac{\not\!q(\not\!p_c+m_c)\Gamma}{2\alpha p_c\cdot q}
   - \frac{\Gamma(\not\!p_b+m_b)\!\not\!q}{2\alpha p_b\cdot q} = 0 \,,
\end{equation}
employing the equations of motion for the heavy quarks. Hence the 
collinear divergence cancels by virtue of the standard Ward identity. 

This completes the proof of the absence of infrared divergences at 
leading power in the hard-scattering kernel for $\bar B_d\to D^+\pi^-$ 
to one-loop order. Similar cancellations are observed at higher orders.
A complete proof of factorization at two-loop order can be found in 
\cite{BBNS00}. Having established that the ``non-factorizable'' diagrams 
of Fig.~\ref{fig6} are dominated by hard gluon exchange (i.e.\ that the 
leading contribution to $J$ arises from $k$ of order $m_b$), we may now
use the fact that $|\vec l_\perp|\ll E$ to expand $A_2$ in powers of
$|\vec l_\perp|/E$. To leading order the expansion simply reduces to 
neglecting $l_\perp$ altogether, which implies $l_q=uq$ and 
$l_{\bar q}=\bar uq$ in (\ref{moms}). As a consequence, we may perform 
the $l_\perp$ integration in (\ref{o8a1a2}) over the pion distribution
amplitude. Defining
\begin{equation}\label{psiphi}
   \int\frac{d^2l_\perp}{16\pi^3}\,
   \frac{\Psi^*(u,\vec l_\perp)}{\sqrt{2 N_c}}
   \equiv \frac{i f_\pi}{4 N_c}\,\Phi_\pi(u) \,, 
\end{equation}
the matrix element of $O_8$ in (\ref{o8a1a2}) becomes
\begin{eqnarray}\label{o8phi}
   \langle D^+\pi^-|O_8|\bar B_d\rangle_{\rm 1\mbox{-}gluon} &=& \\
   &&\hspace*{-4cm}-g_s^2\,\frac{C_F}{8N_c} \int\frac{d^4k}{(2\pi)^4}\,
    \langle D^+|\bar c A_1(k) b|\bar B_d\rangle\,\frac{1}{k^2}\,
    f_\pi\!\int^1_0\!du\,\Phi_\pi(u)\,
    {\rm tr}[\gamma_5\!\not\!q A_2(uq,\bar uq,k)] \,. \nonumber
\end{eqnarray}
On the other hand, putting $y$ on the light-cone in (\ref{piudb}) and 
comparing with (\ref{distamps}), we see that the $l_\perp$-integrated 
wave function $\Phi_\pi(u)$ in (\ref{psiphi}) is precisely the 
light-cone distribution amplitude of the pion. This demonstrates the 
relevance of the light-cone wave function to the factorization formula. 
Note that the collinear approximation for the quark and antiquark 
momenta emerges automatically in the heavy-quark limit. 

After the $k$ integral is performed, the expression (\ref{o8phi}) can 
be cast into the form 
\begin{equation}\label{o8fth}
   \langle D^+\pi^-|O_8|\bar B_d\rangle_{\rm 1\mbox{-}gluon}
   \sim F^{B\to D}(0) \int^1_0\!du\,T_8(u,z)\,\Phi_\pi(u) \,, 
\end{equation}
where $z=m_c/m_b$, $T_8(u,z)$ is the hard-scattering kernel, and 
$F^{B\to D}(0)$ the form factor that parameterizes the 
$\langle D^+|\bar c [\dots] b|\bar B_d\rangle$ matrix element. Because
of the absence of soft and collinear infrared divergences in the gluon 
exchange between the $(\bar cb)$ and $(\bar du)$ currents, the 
hard-scattering kernel $T_8$ is calculable in QCD perturbation theory.

\subsection{Matrix elements at next-to-leading order}
\label{menlo}

We now compute these hard-scattering kernels explicitly to order 
$\alpha_s$. The effective Hamiltonian (\ref{heff18}) can be written as
\begin{eqnarray}\label{bheff}
   {\cal H}_{\rm eff} &=& \frac{G_F}{\sqrt2} V^*_{ud} V_{cb}
    \Bigg\{ \left[ \!\frac{N_c+1}{2N_c}\bar C_+(\mu)
    + \frac{N_c-1}{2N_c}\bar C_-(\mu)
    + \frac{\alpha_s(\mu)}{4\pi}\,\frac{C_F}{2N_c}\,B C_8(\mu)\!
    \right]\!O_0 \nonumber\\
   &&\hspace{2.0cm}\mbox{}+ C_8(\mu)\,O_8 \Bigg\} \,,
\end{eqnarray}
where the scheme-dependent term in the coefficient of the operator 
$O_0$ has been written explicitly. Because the light-quark pair has to 
be in a colour singlet to produce the pion in the leading Fock state, 
only $O_0$ gives a contribution to zeroth order in $\alpha_s$. 
Similarly, to first order in $\alpha_s$ only $O_8$ can contribute. The 
result of evaluating the diagrams in Fig.~\ref{fig6} with an insertion 
of $O_8$ can be presented in a form that holds simultaneously for a 
heavy meson $H=D,D^*$ and a light meson $L=\pi,\rho$, using only that 
the $(\bar u d)$ pair is a colour singlet and that the external quarks 
can be taken on-shell. We obtain ($z=m_c/m_b$)
\begin{eqnarray}\label{delo8}
   \langle H(p') L(q)|O_8|\bar{B}_d(p)\rangle
   &=& \frac{\alpha_s}{4\pi}\frac{C_F}{2N_c}\,i f_L \int_0^1 du\,
   \Phi_L(u) \\ 
   &&\hspace{-4cm}\times 
    \left[ - \left( 6\ln\frac{\mu^2}{m_b^2} + B \right)
    (\langle J_V\rangle - \langle J_A\rangle)
    + F(u,z)\,\langle J_V\rangle - F(u,-z)\,\langle J_A\rangle \right]
    \,, \nonumber
\end{eqnarray}
where 
\begin{equation}\label{qva}
   \langle J_V\rangle
   = \langle H(p')|\bar c \!\not\!q \,b|\bar{B}_d(p)\rangle , \qquad
   \langle J_A\rangle
   = \langle H(p')|\bar c\!\not\!q\gamma_5 b\,|\bar{B}_d(p)\rangle \,.
\end{equation}
It is worth noting that even after computing the one-loop correction 
the $(\bar u d)$ pair retains its $V-A$ structure. This, together with 
(\ref{distamps}), implies that the form of (\ref{delo8}) is identical 
for pions and longitudinally polarized $\rho$ mesons. (The production of  
transversely polarized $\rho$ mesons is power suppressed in 
$\Lambda_{\rm QCD}/m_b$.) The function $F(u,z)$ appearing in 
(\ref{delo8}) is given by
\begin{equation}\label{ff}
   F(u,z) = \left( 3 + 2 \ln\frac{u}{\bar u} \right) \ln z^2 - 7
   + f(u,z) + f(\bar u,1/z) \,,
\end{equation}
where 
\begin{eqnarray}\label{fxz}
   f(u,z) &=& - \frac{u(1-z^2)[3(1-u (1-z^2))+z]}{[1-u(1-z^2)]^2}
    \ln[u(1-z^2)] - \frac{z}{1-u(1-z^2)} \nonumber\\
   &&\hspace{-1.6cm}\mbox{}+ 2 \left[ \frac{\ln[u(1-z^2)]}{1-u(1-z^2)}
    - \ln^2[u(1-z^2)] - \mbox{Li}_2[1-u(1-z^2)] - \{ u\to\bar u \}
    \right] \,,
\end{eqnarray}
and $\mbox{Li}_2(x)$ is the dilogarithm. The contribution of $f(u,z)$ 
in (\ref{ff}) comes from the first two diagrams in Fig.~\ref{fig6} with 
the gluon coupling to the $b$ quark, whereas $f(\bar u,1/z)$ arises 
from the last two diagrams with the gluon coupling to the charm quark. 
Note that the terms in the large square brackets in the definition of 
the function $f(u,z)$ vanish for a symmetric light-cone 
distribution amplitude. These terms can be dropped if the light 
final-state meson is a pion or a $\rho$ meson, but they are relevant, 
e.g., for the discussion of Cabibbo-suppressed decays such as 
$\bar B_d\to D^{(*)+} K^-$ and $\bar B_d\to D^{(*)+} K^{*-}$. 

The discontinuity of the amplitude, which is responsible for the 
occurrence of the strong rescattering phase, arises from $f(\bar u,1/z)$ 
and can be obtained by recalling that $z^2$ is $z^2-i\epsilon$ with 
$\epsilon>0$ infinitesimal. We find 
\begin{eqnarray}
   \frac{1}{\pi}\,\mbox{Im}\,F(u,z)
   &=& - \frac{(1-u)(1-z^2)[3(1-u (1-z^2))+z]}{[1-u(1-z^2)]^2}
    \nonumber\\
   &&\hspace{-2.2cm}
    \mbox{}- 2 \left[ \ln[1-u(1-z^2)] + 2\ln u
    + \frac{z^2}{1-u(1-z^2)} - \{ u\to\bar u \} \right] \,.
\end{eqnarray}

As mentioned above, (\ref{delo8}) is applicable to all decays of the 
type $\bar B_d\to D^{(*)+}L^-$, where $L$ is a light hadron such as a 
pion or a (longitudinally polarized) $\rho$ meson. Only the operator 
$J_V$ contributes to $\bar B_d\to D^+ L^-$, and only $J_A$ contributes 
to $\bar B_d\to D^{*+} L^-$. Our result can therefore be written as 
\begin{equation}\label{bdpi18}
   \langle D^+ L^-|O_{0,8}|\bar B_d\rangle
   = \langle D^+|\bar c\gamma^\mu(1-\gamma_5)b|\bar B_d\rangle
   \cdot i f_L q_\mu \int^1_0\!du\,T_{0,8}(u,z)\,\Phi_L(u) \,,
\end{equation}
where $L=\pi$, $\rho$, and the hard-scattering kernels are
\begin{eqnarray}\label{t1uz}
   T_0(u,z) &=& 1 + O(\alpha^2_s) \,, \nonumber\\
   T_8(u,z) &=& \frac{\alpha_s}{4\pi}\,\frac{C_F}{2N_c} \left[
    - 6\ln\frac{\mu^2}{m_b^2} - B + F(u,z) \right] + O(\alpha^2_s) \,.
\end{eqnarray}
When the $D$ meson is replaced by a $D^*$ meson, the result is identical 
except that $F(u,z)$ must be replaced with $F(u,-z)$. Since no 
order-$\alpha_s$ corrections exist for $O_0$, the matrix element retains 
its leading-order factorized form
\begin{equation}\label{o1me}
   \langle D^+L^-|O_0|\bar B_d\rangle = i f_L q_\mu\,
   \langle D^+|\bar c\gamma^\mu(1-\gamma_5)b|\bar B_d\rangle
\end{equation}
to this accuracy. From (\ref{fxz}) it follows that $T_8(u,z)$ tends to 
a constant as $u$ approaches the endpoints ($u\to 0$, $1$). (This is 
strictly true for the part of $T_8(u,z)$ that is symmetric in 
$u\leftrightarrow\bar u$; the asymmetric part diverges logarithmically 
as $u\to 0$, which however does not affect the power behaviour and the 
convergence properties in the endpoint region.) Therefore the 
contribution to (\ref{bdpi18}) from the endpoint region is suppressed, 
both by phase space and by the endpoint suppression intrinsic to 
$\Phi_L(u)$. Consequently, the emitted light meson is indeed dominated 
by energetic constituents, as required for the self-consistency of the 
factorization formula.  

The final result for the class-I, non-leptonic 
$\bar B_d\to D^{(*)+} L^-$ decay amplitudes, in the heavy-quark limit 
and at next-to-leading order in $\alpha_s$, can be compactly expressed 
in terms of the matrix elements of a ``transition operator''
\begin{equation}\label{heffa1}
   {\cal T} = \frac{G_F}{\sqrt 2} V^*_{ud}V_{cb}
   \Big[ a_1(D L)\,Q_V - a_1(D^* L)\,Q_A \Big] \,,
\end{equation}
where
\begin{equation}\label{qva2}
   Q_V = \bar c\gamma^\mu b\,\otimes\,\bar d\gamma_\mu(1-\gamma_5)u \,,
   \qquad
   Q_A = \bar c\gamma^\mu\gamma_5 b\,\otimes\, 
   \bar d\gamma_\mu(1-\gamma_5)u \,,
\end{equation}
and hadronic matrix elements of $Q_{V,A}$ are understood to be evaluated
in factorized form, i.e.\ 
\begin{equation}
   \langle D L|j_1\otimes j_2|\bar B\rangle
   \equiv \langle D|j_1|\bar B\rangle\,\langle L|j_2|0\rangle \,.
\end{equation} 
Eq.~(\ref{heffa1}) defines the quantities $a_1(D^{(*)} L)$, which 
include the leading ``non-factorizable'' corrections, in a 
renormalization-scale and -scheme independent way. To leading power in 
$\Lambda_{\rm QCD}/m_b$ these quantities should not be interpreted as 
phenomenological parameters (as is usually done), because they are 
dominated by hard gluon exchange and thus calculable in QCD. At 
next-to-leading order we get
\begin{eqnarray}\label{a1dpi}
   a_1(D L) &=& \frac{N_c+1}{2N_c}\bar C_+(\mu)
    + \frac{N_c-1}{2N_c}\bar C_-(\mu) \nonumber\\
   &&\mbox{}+ \frac{\alpha_s}{4\pi}\frac{C_F}{2N_c}\,C_8(\mu) \left[
    - 6\ln\frac{\mu^2}{m_b^2} + \int^1_0 du\,F(u,z)\,\Phi_L(u)
    \right] \,, \nonumber\\
   a_1(D^* L) &=& \frac{N_c+1}{2N_c}\bar C_+(\mu)
    + \frac{N_c-1}{2N_c}\bar C_-(\mu) \nonumber\\
   &&\mbox{}+ \frac{\alpha_s}{4\pi}\frac{C_F}{2N_c}\,C_8(\mu) \left[
    - 6\ln\frac{\mu^2}{m_b^2} + \int^1_0 du\,F(u,-z)\,\Phi_L(u) \right]
    \,.
\end{eqnarray}
We observe that the scheme-dependent terms parameterized by $B$ have
canceled between the coefficient of $O_0$ in (\ref{bheff}) and the 
matrix element of $O_8$ in (\ref{bdpi18}). Likewise, the $\mu$ 
dependence of the terms in brackets in (\ref{a1dpi}) cancels against 
the scale dependence of the coefficients $\bar C_\pm(\mu)$, ensuring a 
consistent result at next-to-leading order. 
The coefficients $a_1(D L)$ and $a_1(D^* L)$ are seen to be 
non-universal, i.e.\ they depend explicitly on the nature of the 
final-state mesons. This dependence enters via the light-cone 
distribution amplitude of the light emission meson and via the analytic 
form of the hard-scattering kernel ($F(u,z)$ vs.\ $F(u,-z)$). However, 
the non-universality enters only at next-to-leading order.

Using the fact that violations of heavy-quark spin symmetry require 
hard gluon exchange, Politzer and Wise have computed the 
``non-factorizable'' vertex corrections to the decay-rate ratio of the 
$D\pi$ and $D^*\pi$ final states many years ago \cite{PW91}. In the
context of our formalism, this calculation requires the symmetric part 
(with respect to $u\leftrightarrow\bar u$) of the difference 
$F(u,z)-F(u,-z)$. Explicitly,
\begin{equation}\label{dpdsp}
   \frac{\Gamma(\bar B_d\to D^+\pi^-)}{\Gamma(\bar B_d\to D^{*+}\pi^-)}
   = \left|
   \frac{\langle D^+|\bar c\!\not\! q(1-\gamma_5)b|\bar B_d\rangle}
        {\langle D^{*+}|\bar c\!\not\! q(1-\gamma_5)b|\bar B_d\rangle}
   \right|^2 \left| \frac{a_1(D\pi)}{a_1(D^*\pi)} \right|^2 \,,
\end{equation}
where for simplicity we neglect the light meson masses as well as the
mass difference between $D$ and $D^*$ in the phase-space for the two 
decays. At next-to-leading order 
\begin{equation}
   \left| \frac{a_1(D\pi)}{a_1(D^*\pi)} \right|^2
   = 1 + \frac{\alpha_s}{4\pi}\frac{C_F}{N_c}\frac{C_8}{C_0}\,
   {\rm Re}\int^1_0 du \left[ F(u,z)-F(u,-z) \right]\,\Phi_\pi(u) \,.
\end{equation}
Our result for the symmetric part of $F(u,z)-F(u,-z)$ coincides with 
that found in \cite{PW91}.

\section{Power-suppressed contributions}
\label{sec:power}

Up to this point we have presented arguments in favour of factorization
of non-leptonic $B$-decay amplitudes in the heavy-quark limit, and have
explored in detail how the factorization formula works at one-loop 
order for the decays $\bar B_d\to D^{(*)+} L^-$. It is now time to 
show that other contributions not considered so far are indeed power
suppressed. This is necessary to fully establish the factorization 
formula. Besides, it will also provide some numerical estimates of the 
corrections to the heavy-quark limit.

We start by discussing interactions involving the spectator quark and 
weak annihilation contributions, before turning to the more delicate 
question of the importance of non-valence Fock states.

\subsection{Interactions with the spectator quark}
\label{subsec:hardspec}

\begin{figure}[t]
\vspace{-3.4cm}
\epsfxsize=18cm
\centerline{\epsffile{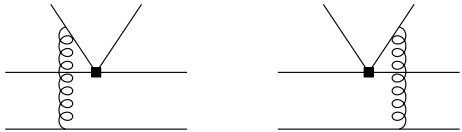}}
\vspace{-19.9cm}
\caption{\label{fig8}
``Non-factorizable'' spectator interactions.}
\end{figure}

Clearly, the diagrams shown in Fig.~\ref{fig8} cannot be associated 
with the form-factor term in the factorization formula (\ref{fff}). 
We will now show that for $B$ decays into a heavy-light final state 
their contribution is power suppressed in the heavy-quark limit. 
(This suppression does {\em not\/} occur for decays into two 
light mesons, where hard spectator interactions contribute at leading 
power. In this case, they contribute to the kernels $T_i^{II}$ in the 
factorization formula (second term in Fig.~\ref{fig1}).)

In general, ``non-factorizable'' diagrams involving an interaction with 
the spectator quark would impede factorization if there existed a soft 
contribution at leading power. While such terms are present in each of 
the two diagrams separately, they cancel in the sum over the two gluon 
attachments to the $(\bar u d)$ pair by virtue of the same 
colour-transparency argument that was applied to the 
``non-factorizable'' vertex corrections. 

Focusing again on decays into a heavy and a light meson, such as 
$\bar B_d\to D^+\pi^-$, we still need to show that the contribution 
remaining after the soft cancellation is power suppressed 
relative to the leading-order contribution (\ref{abd}). A 
straightforward calculation leads to the following (simplified) result 
for the sum of the two diagrams:
\begin{eqnarray}\label{specd}
   {\cal A}(\bar B_d\to D^+\pi^-)_{\rm spec} 
   &\sim& G_F\,f_\pi f_D f_B\,\alpha_s \nonumber\\
   &&\times \int_0^1\!\frac{d\xi}{\xi}\,\Phi_B(\xi)
    \int_0^1\!\frac{d\eta}{\eta}\,\Phi_D(\eta)
    \int_0^1\!\frac{d u}{u}\,\Phi_{\pi}(u) \nonumber\\
   &\sim& G_F\,\alpha_s\,m_b\,\Lambda_{\rm QCD}^2 \,.
\end{eqnarray}
This is indeed power suppressed relative to (\ref{abd}). Note that 
the gluon virtuality is of order $\xi\eta\,m_b^2\sim\Lambda_{\rm QCD}^2$ 
and so, strictly speaking, the calculation in terms of light-cone 
distribution amplitudes cannot be justified. Nevertheless, we use 
(\ref{specd}) to deduce the scaling behaviour of the soft contribution, 
as we did for the heavy-light form factor in Sect.~\ref{formfactor}.

\subsection{Annihilation topologies}
\label{subsec:annihilation}

\begin{figure}[t]
\vspace{-4.3cm}
\epsfxsize=15cm
\centerline{\epsffile{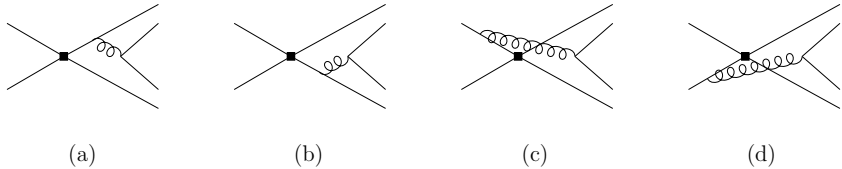}}
\vspace{-14.5cm}
\caption{\label{fig9}
Annihilation diagrams.}
\end{figure}

Our next concern are the annihilation diagrams shown in Fig.~\ref{fig9},
which also contribute to the decay $\bar B_d\to D^+ \pi^-$. The hard 
part of these diagrams could, in principle, be absorbed into 
hard-scattering kernels of the type $T_i^{II}$. The soft part, if 
unsuppressed, would violate factorization. However, we will see that
the hard part as well as the soft part are suppressed by at least one 
power of $\Lambda_{\rm QCD}/m_b$.

The argument goes as follows. We write the annihilation amplitude as 
\begin{eqnarray}\label{ann2}
   {\cal A}(\bar{B}_d\to D^+ \pi^-)_{\rm ann} 
   &\sim& G_F\,f_\pi f_D f_B\,\alpha_s \nonumber\\
   &\times& \int_0^1\! d\xi\,d\eta\,du\,\Phi_B(\xi)\,\Phi_D(\eta)\,
    \Phi_\pi(u)\,T^{\rm ann}(\xi,\eta,u) \,,~~
\end{eqnarray}
where the dimensionless function $T^{\rm ann}(\xi,\eta,u)$ is a product 
of propagators and vertices. The product of decay constants scales as 
$\Lambda_{\rm QCD}^4/m_b$. Since $d\xi\,\Phi_B(\xi)$ scales as 1 and so 
does $d\eta\,\Phi_D(\eta)$, while $du\,\Phi_\pi(u)$ is never larger than 
1, the amplitude can only compete with the leading-order result 
(\ref{abd}) if $T^{\rm ann}(\xi,\eta,u)$ can be made of order 
$(m_b/\Lambda_{\rm QCD})^3$ or larger. Since $T^{\rm ann}(\xi,\eta,u)$ 
contains only two propagators, this can be achieved only if both quarks 
the gluon splits into are soft, in which case 
$T^{\rm ann}(\xi,\eta,u)\sim (m_b/\Lambda_{\rm QCD})^4$. But then 
$du\,\Phi_\pi(u)\sim(\Lambda_{\rm QCD}/m_b)^2$, so that this 
contribution is power suppressed.

\subsection{Non-leading Fock states}
\label{otherfock}

Our discussion so far concentrated on contributions related to the 
quark--antiquark components of the meson wave functions. We now present 
qualitative arguments that justify this restriction to the 
valence-quark Fock components. Some of these arguments are standard 
\cite{LB80,EfRa80}. We will argue that higher Fock states yield only 
subleading contributions in the heavy-quark limit.

\begin{figure}[t]
\vspace{-3.3cm}
\epsfxsize=18cm
\centerline{\epsffile{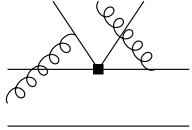}}
\vspace{-20.0cm}
\caption{\label{fig10}
Diagram that contributes to the hard-scattering kernel involving a 
quark--antiquark--gluon distribution amplitude of the $B$ meson and the 
emitted light meson.}
\end{figure}

\subsubsection{Additional hard partons}

An example of a diagram that would contribute to a hard-scattering 
function involving quark--antiquark--gluon components of the emitted 
meson and the $B$ meson is shown in Fig.~\ref{fig10}. For light mesons, 
higher Fock components are related to higher-order terms in the 
collinear expansion, including the effects of intrinsic transverse 
momentum and off-shellness of the partons by gauge invariance. The 
assumption is that the additional partons are collinear and carry a 
finite fraction of the meson momentum in the heavy-quark limit. Under 
this assumption, it is easy to see that adding additional partons to 
the Fock state increases the number of off-shell propagators in a given 
diagram (compare Fig.~\ref{fig10} to Fig.~\ref{fig4}). This implies 
power suppression in the heavy-quark expansion. Additional partons 
in the $B$-meson wave function are always soft, as is the spectator 
quark. Nevertheless, when these partons are connected to the 
hard-scattering amplitudes the virtuality of the additional propagators 
is still of order $m_b\Lambda_{\rm QCD}$, which is sufficient to 
guarantee power suppression.

\begin{figure}[h]
\epsfxsize=7cm
\centerline{\epsffile{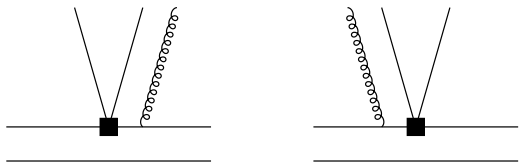}}
\caption{\label{fig:qqg}
The contribution of the $q\bar q g$ Fock state to the 
$\bar B_d\to D^+\pi^-$ decay amplitude.}
\end{figure}

Let us study in more detail how the power suppression arises for the 
simplest non-trivial example, where the pion is composed of a quark, an 
antiquark, and an additional gluon. The contribution of this 3-particle 
Fock state to the $B\to D\pi$ decay amplitude is shown in 
Fig.~\ref{fig:qqg}. It is convenient to use the Fock--Schwinger gauge, 
which allows us to express the gluon field $A_\lambda$ in terms of the 
field-strength tensor $G_{\rho\lambda}$ via
\begin{equation}\label{axg}
   A_\lambda(x) = \int^1_0\!dv\,v x^\rho\,G_{\rho\lambda}(vx) \,.
\end{equation}
Up to twist-4 level, there are three quark--antiquark--gluon matrix 
elements that could potentially contribute to the diagrams shown in 
Fig.~\ref{fig:qqg}. Due to the $V-A$ structure of the weak-interaction 
vertex, the only relevant three-particle light-cone wave function has 
twist-4 and is given by \cite{Khod98a,BF90}
\begin{eqnarray}\label{pidgu}
   &&\langle\pi(q)|\bar d(0)\gamma_\mu\gamma_5\,
    g_s G_{\alpha\beta}(vx)\,u(0)|0\rangle \nonumber\\
   &=& f_\pi(q_\beta g_{\alpha\mu} - q_\alpha g_{\beta\mu})
    \int{\cal D}u\,\phi_\perp(u_i)\,e^{iv u_3 q\cdot x} \nonumber\\
   &+& f_\pi\,\frac{q_\mu}{q\cdot x}\,
    (q_\alpha x_\beta - q_\beta x_\alpha)
    \int{\cal D}u\,\left( \phi_\perp(u_i) + \phi_\parallel(u_i) \right)
    e^{iv u_3 q\cdot x} \,.
\end{eqnarray}
Here $\int{\cal D}u\equiv\int^1_0\!du_1\,du_2\,du_3\,
\delta(1-u_1-u_2-u_3)$, with $u_1$, $u_2$ and $u_3$ the fractions of the 
pion momentum carried by the quark, antiquark and gluon, respectively. 
Evaluating the diagrams in Fig.~\ref{fig:qqg}, and neglecting the 
charm-quark mass for simplicity, we find
\begin{equation}\label{o8qqg}
   \langle D^+\pi^-|O_8|\bar B_d\rangle_{q\bar q g}
   = i f_\pi\,\langle D^+|\bar c\!\not\!q(1-\gamma_5)b|\bar B_d\rangle
   \int{\cal D}u\,\frac{2\phi_\parallel(u_i)}{u_3\, m^2_b} \,.
\end{equation}
Since $\phi_\parallel\sim\Lambda^2_{\rm QCD}$, the suppression by two
powers of $\Lambda_{\rm QCD}/m_b$ compared to the leading-order matrix 
element is obvious. Note that due to G-parity $\phi_\parallel$ is 
antisymmetric in $u_1\leftrightarrow u_2$ for a pion, so that 
(\ref{o8qqg}) vanishes in this case.

\subsubsection{Additional soft partons}

\begin{figure}[t]
\vspace{-3.4cm}
\epsfxsize=18.2cm
\centerline{\epsffile{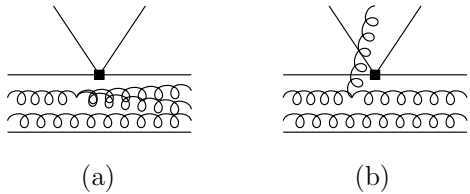}}
\vspace{-19.3cm}
\caption{\label{fig11}
(a) Soft overlap contribution which is part of the $B\to D$ form factor. 
(b) Soft overlap with the pion which would violate factorization, if it 
were unsuppressed.}
\end{figure}

A more precarious situation may arise when the additional Fock 
components carry only a small fraction of the meson momentum, contrary 
to the assumption made above. It is usually argued \cite{LB80,EfRa80} 
that these configurations are suppressed, because they occupy only 
a small fraction of the available phase space (since $\int du_i\sim 
\Lambda_{\rm QCD}/m_b$ when the parton that carries momentum fraction 
$u_i$ is soft). This argument does not apply when the process involves 
heavy mesons. Consider, for example, the diagram shown in 
Fig.~\ref{fig11} (a) for the decay $B\to D\pi$. Its contribution 
involves the overlap of the $B$-meson wave function involving additional 
soft gluons with the wave function of the $D$ meson, also containing 
soft gluons. There is no reason to suppose that this overlap is 
suppressed relative to the soft overlap of the valence-quark wave 
functions. It represents (part of) the overlap of the ``soft cloud'' 
around the $b$ quark with (part of) the ``soft cloud'' around the $c$ 
quark after the weak decay. The partonic decomposition of this cloud is 
unrestricted up to global quantum numbers. (In the case where the $B$ 
meson decays into two light mesons, there is a form-factor suppression 
$\sim(\Lambda_{\rm QCD}/m_b)^{3/2}$ for the overlap of the 
valence-quark wave functions, but once this price is paid there is again 
no reason for further suppression of additional soft gluons in the 
overlap of the $B$-meson wave function and the wave function of the 
recoiling meson.)

The previous paragraph essentially repeated our earlier argument against 
the hard-scattering approach, and in favour of using the $B\to D$ form 
factor as an input to the factorization formula. However, given the 
presence of additional soft partons in the $B\to D$ transition, we must 
now argue that it is unlikely that the emitted pion drags with it one of 
these soft partons, for instance a soft gluon that goes into the pion 
wave function, as shown in Fig.~\ref{fig11} (b). Notice that if the 
$(q\bar q)$ pair is produced in a colour-octet state, at least one gluon 
(or a further $(q\bar q)$ pair) must be pulled into the emitted meson if 
the decay is to result in a two-body final state. What suppresses the 
process shown in Fig.~\ref{fig11} (b) relative to the one in 
Fig.~\ref{fig11} (a) even if the emitted $(q\bar q)$ pair is in a 
colour-octet state? 

\begin{figure}[t]
\vspace{-4.1cm}
\epsfxsize=18cm
\centerline{\epsffile{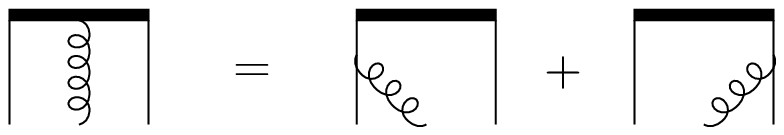}}
\vspace{-20.1cm}
\caption{\label{fig12}
Quark--antiquark--gluon distribution amplitude in the gluon endpoint 
region.}
\end{figure}

It is once more colour transparency that saves us. The dominant 
configuration has both quarks carry a large fraction of the pion 
momentum, and only the gluon might be soft. In this situation we can 
apply a non-local ``operator product expansion'' to determine the 
coupling of the soft gluon to the small $(q\bar q)$ pair \cite{BBNS00}. 
The gluon endpoint behaviour of the $q\bar{q}g$ wave function is then 
determined by the sum of the two diagrams shown on the right-hand side 
in Fig.~\ref{fig12}. The leading term (for small gluon 
momentum) cancels in the sum of the two diagrams, because the meson 
(represented by the black bar) is a colour singlet. This cancellation, 
which is exactly the same cancellation needed to demonstrate that 
``non-factorizable'' vertex corrections are dominated by hard gluons, 
provides one factor of $\Lambda_{\rm QCD}/m_b$ needed to show that 
Fig.~\ref{fig11} (b) is power suppressed relative to Fig.~\ref{fig11} 
(a). 

In summary, we have (qualitatively) covered all possibilities for 
non-valence contributions to the decay amplitude and find that they 
are all power suppressed in the heavy-quark limit.

\section{Limitations of the factorization approach}
\label{limitations}

The factorization formula (\ref{fff}) holds in the heavy-quark limit 
$m_b\to\infty$. Corrections to the asymptotic limit are power-suppressed 
in the ratio $\Lambda_{\rm QCD}/m_b$ and, generally speaking, do not 
assume a factorized form. Since $m_b$ is fixed to about 5\,GeV in the 
real world, one may worry about the magnitude of power corrections to
hadronic $B$-decay amplitudes. Naive dimensional analysis would suggest
that these corrections should be of order 10\% or so. We now discuss 
several reasons why some power corrections could turn out to be 
numerically larger than suggested by the parametric suppression factor 
$\Lambda_{\rm QCD}/m_b$. Most of these ``dangerous'' corrections occur 
in more complicated, rare hadronic $B$ decays into two light mesons, 
but are absent in decays such as $B\to D\pi$.

\subsection{Several small parameters}

Large non-factorizable power corrections may arise if the leading-power, 
factorizable term is somehow suppressed. There are several possibilities 
for such a suppression, given a variety of small parameters that may 
enter into the non-leptonic decay amplitudes.
\begin{itemize}
\item[i)]
The hard, ``non-factorizable'' effects computed using the factorization 
formula occur at order $\alpha_s$. Some other interesting effects such 
as final-state interactions appear first at this order. For instance,
strong-interaction phases due to hard interactions are of order 
$\alpha_s$, while soft rescattering phases are of order 
$\Lambda_{\rm QCD}/m_b$. Since for realistic $B$ mesons $\alpha_s$ is 
not particularly large compared to $\Lambda_{\rm QCD}/m_b$, we should 
not expect that these phases can be calculated with great precision. In 
practice, however, it is probably more important to know that the 
strong-interaction phases are parametrically suppressed in the 
heavy-quark limit and thus should be small. (This does not apply if the 
real part of the decay amplitudes is suppressed for some reason; see 
below.)
\item[ii)]
If the leading, lowest-order (in $\alpha_s$) contribution to the decay
amplitude is colour suppressed, as occurs for the class-II decay 
$\bar B_d\to\pi^0\pi^0$, then perturbative and power corrections can be 
sizeable. In such a case even the hard strong-interaction phase of the
amplitude can be large \cite{BBNS99,BBNS00}. But at the same time soft 
contributions could be potentially important, so that in some cases only 
an order-of-magnitude estimate of the amplitude may be possible.
\item[iii)]
The effective Hamiltonian (\ref{effham}) contains many Wilson 
coefficients $C_i$ that are small relative to $C_1\approx 1$. 
There are decays for which the entire leading-power contribution is 
suppressed by small Wilson coefficients, but some power-suppressed 
effects are not. An example of this type is $B^-\to K^- K^0$. This decay 
proceeds through a penguin operator $b\to d s\bar s$ at leading power. 
But the annihilation contribution, which is power suppressed, can occur 
through the current--current operator with large Wilson coefficient 
$C_1$. Our approach does not apply to such (presumably) 
annihilation-dominated decays, unless a systematic treatment of 
annihilation amplitudes can be found.
\item[iv)]
Some amplitudes may be suppressed by a combination of small CKM matrix 
elements. For example, $B\to\pi K$ decays receive large penguin 
contributions despite their small Wilson coefficients, because the 
so-called tree amplitude is CKM suppressed. This is not a problem for 
factorization, since it applies to the penguin and the tree amplitudes. 
We are not aware of any case (for ordinary $B$ mesons) in which a 
purely power-suppressed term is CKM enhanced and which would therefore 
dominate the decay amplitude. (But this situation could occur for 
$B^-_c\to\bar D^0 K^-$, where the QCD dynamics is similar if we consider 
the charm quark as a light quark.)
\end{itemize}

\subsection{Power corrections enhanced by small quark masses}
\label{chiral}

There is another enhancement of power-suppressed effects for some decays 
into two light mesons, connected with the curious numerical fact that 
\begin{equation}\label{mupi}
   2\mu_\pi\equiv \frac{2 m_\pi^2}{m_u+m_d}
   = -\frac{4\langle\bar{q}q\rangle}{f_\pi^2} \approx 3\,\mbox{GeV}
\end{equation}
is much larger than its naive scaling estimate $\Lambda_{\rm QCD}$. 
(Here $\langle\bar q q\rangle=\langle 0|\bar u u|0\rangle
=\langle 0|\bar d d|0\rangle$ is the quark condensate.) Consider the 
contribution of the penguin operator 
$O_6=(\bar d_i b_j)_{V-A} (\bar u_j u_i)_{V+A}$ to the 
$\bar B_d\to\pi^+\pi^-$ decay amplitude. The leading-order graph of 
Fig.~\ref{fig4} results in the expression
\begin{equation}\label{lotw3}
   \langle\pi^+\pi^-|(\bar d_i b_j)_{V-A} (\bar u_j u_i)_{V+A}
   |\bar B_d\rangle = i m_B^2\,F_+^{B\to \pi}(0)\,f_\pi\, 
   \times\frac{2\mu_\pi}{m_b} \,,
\end{equation}
which is formally a $\Lambda_{\rm QCD}/m_b$ power correction compared 
to the corresponding matrix element of a product of two left-handed 
currents, but numerically large due to (\ref{mupi}). We would not have 
to worry about such terms if they could all be identified and the 
factorization formula (\ref{fff}) applied to them, since in this case 
higher-order perturbative corrections would not contain non-factorizing 
infrared logarithms. However, this is not the case.

After including radiative corrections, the matrix element on the 
left-hand side of (\ref{lotw3}) is expressed as a non-trivial 
convolution with pion light-cone distribution amplitudes. The terms 
involving $\mu_\pi$ can be related to two-particle twist-3 (rather than 
leading twist-2) distribution amplitudes, conventionally called 
$\Phi_p(u)$ and $\Phi_\sigma(u)$. We find that the radiative
corrections to the matrix element in (\ref{lotw3}) do indeed factorize.
However, at the same order there appear twist-3 corrections to the hard 
spectator interaction shown in Fig.~\ref{fig8}, and these contributions
contain an endpoint divergence (related to the fact that the 
distribution amplitudes $\Phi_p(u)$ and $\Phi'_\sigma(u)$ do not vanish 
at the endpoints). In other words, the twist-3 ``corrections'' to the 
hard spectator term in the second factorization formula in (\ref{fff}) 
relative to the ``leading'' twist-2 contributions are of the form 
$\alpha_s\times\,$logarithmic divergence, which we interpret as being 
of order 1. The non-factorizing character of the ``chirally-enhanced'' 
power corrections can introduce a substantial uncertainty in some decay 
modes \cite{BBNSfuture}. As in the related situation for the pion form 
factor \cite{GT82}, one may argue that the endpoint divergence is 
suppressed by a Sudakov form factor. However, it is likely that when 
$m_b$ is not large enough to suppress these chirally-enhanced terms, 
then it is also not large enough to make Sudakov suppression effective. 

We stress that the chirally-enhanced terms do not appear in decays into 
a heavy and a light meson such as $B\to D\pi$, because these decays 
have no penguin contribution and no contribution from the hard 
spectator interaction. Hence, the twist-3 light-cone distribution 
amplitudes responsible for chirally-enhanced power corrections do not 
enter in the evaluation of the decay amplitudes.

\subsection{Non-leptonic decays when $M_2$ is not light}
\label{hl}

The analysis of non-leptonic decay amplitudes in Sect.~\ref{nlamp} 
referred to decays where the emission particle $M_2$ is a light meson. 
We now briefly discuss the case where $M_2$ is heavy. 

Suppose that $M_2$ is a $D$ meson, whereas the meson that picks up the 
spectator quark can be heavy or light. Examples of this type are the 
decays $\bar B_d\to\pi^0 D^0$ and $\bar B_d\to D^+ D^-$. It is 
intuitively clear that factorization must be problematic in these cases, 
because the heavy $D$ meson has a large overlap with the $\bar B\pi$ or 
$\bar B D$ systems, which are dominated by soft processes. In more 
detail, we consider the coupling of a gluon to the two quarks that form 
the emitted $D$ meson, i.e.\ the pairs of diagrams in Fig.~\ref{fig6} 
(a+b), (c+d) and Fig.~\ref{fig8}. Denoting the gluon momentum by $k$, 
the quark momenta by $l_q$ and $l_{\bar{q}}$, and the $D$-meson momentum 
by $q$, we find that the gluon couples to the ``current''
\begin{equation}\label{jj}
   J_\lambda
   = \frac{\gamma_\lambda(\not\!l_q+\not\!k+m_c)\Gamma}
          {2 l_q\cdot k+k^2}
   - \frac{\Gamma(\not\!l_{\bar q}+\not\!k)\gamma_\lambda}
          {2 l_{\bar q}\cdot k+k^2} \,,
\end{equation}
where $\Gamma$ is part of the weak decay vertex. When $k$ is soft (all 
components of order $\Lambda_{\rm QCD}$), each of the two terms scales 
as $1/\Lambda_{\rm QCD}$. Taking into account the complete amplitude as 
done explicitly in Sect.~\ref{oneloopcancel}, we can see that the 
decoupling of soft gluons requires that the two terms in (\ref{jj}) 
cancel, leaving a remainder of order $1/m_b$. This cancellation does 
indeed occur when $M_2$ is a light meson, since in this case $l_q$ and 
$l_{\bar{q}}$ are dominated by their longitudinal components. When 
$M_2$ is heavy, the momenta $l_q$ and $l_{\bar q}$ are asymmetric, with 
all components of the light antiquark momentum $l_{\bar q}$ of order 
$\Lambda_{\rm QCD}$ in the $B$- or $D$-meson rest frames, while the 
zero-component of $l_q$ is of order $m_b$. Hence the current 
can be approximated by 
\begin{equation}
   J_\lambda \approx \frac{\delta_{\lambda 0}\Gamma}{k_0}
   - \frac{\Gamma(\not\!l_{\bar q}+\not\!k)\gamma_\lambda}
          {2 l_{\bar q}\cdot k+k^2}
   \sim \frac{1}{\Lambda_{\rm QCD}} \,,
\end{equation}
and the soft cancellation does not occur. (The on-shell condition for 
the charm quark has been used to arrive at this equation.)

It follows that the emitted $D$ meson does not factorize from the rest 
of the process, and that a factorization formula analogous to (\ref{fff}) 
does not apply to decays such as $\bar B_d\to\pi^0 D^0$ and 
$\bar B_d\to D^+ D^-$. An important implication of this statement is 
that one should also not expect naive factorization to work in these 
cases. In other words, we expect that ``non-factorizable'' corrections 
modify the factorized decay amplitudes by terms of order 1. 

\subsection{Difficulties with charm}
\label{subsec:diffcharm}

There are decay modes, such as $B^-\to D^0\pi^-$, in which the 
spectator quark can go to either of the two final-state mesons. The 
factorization formula (\ref{fff}) applies to the contribution that 
arises when the spectator quark goes to the $D$ meson, but not when 
the spectator quark goes to the pion. However, even in the latter case 
we may use naive factorization to estimate the power behaviour of the 
decay amplitude. Adapting (\ref{abd}) to the decay $B^-\to D^0\pi^-$, 
we find that the non-factorizing (class-II) amplitude is suppressed 
compared to the factorizing (class-I) amplitude by
\begin{equation}\label{onetwo}
   \frac{{\cal A}(B^-\to D^0\pi^-)_{\rm class\mbox{-}II}}
        {{\cal A}(B^-\to D^0\pi^-)_{\rm class\mbox{-}I}}
   \sim \frac{F^{B\to \pi}(m_D^2)\,f_D}{F^{B\to D}(m_\pi^2)\,f_\pi} 
   \sim \left( \frac{\Lambda_{\rm QCD}}{m_b} \right)^2 .
\end{equation}
Here we use that $F^{B\to\pi}(q^2)\sim(\Lambda_{\rm QCD}/m_b)^{3/2}$ 
even for $q^2\sim m_b^2$, as long as $q_{max}^2-q^2$ is also of order 
$m_b^2$. (It follows from our definition of heavy final-state mesons 
that these conditions are fulfilled.) As a consequence, strictly speaking 
factorization {\em does\/} hold for $B^-\to D^0\pi^-$ decays in the 
sense that the class-II contribution is power suppressed with respect
to the class-I contribution. 

Unfortunately, the scaling behaviour for real $B$ and $D$ mesons is far 
from the estimate (\ref{onetwo}) valid in the heavy-quark limit. Based 
on the dominance of the class-I amplitude we would expect that
\begin{equation}
   R = 
   \frac{\mbox{Br}(B^-\to D^0\pi^-)}{\mbox{Br}(\bar{B}_d\to D^+\pi^-)}
   \approx 1
\end{equation}
in the heavy-quark limit. This contradicts existing data which yield 
$R=1.89\pm 0.35$, despite the additional colour suppression of the 
class-II amplitude. One reason for the failure of power counting lies 
in the departure of the decay constants and form factors from naive 
power counting. The following compares the power counting to the actual 
numbers (square brackets):
\begin{equation}\label{scalingviolations}
   \frac{f_D}{f_\pi} \sim \left( \frac{\Lambda_{\rm QCD}}{m_c}
   \right)^{1/2} \,\,[\approx 1.5] \,,
   \qquad
   \frac{F_+^{B\to \pi}(m_D^2)}{F_+^{B\to D}(m_\pi^2)}
   \sim \left( \frac{\Lambda_{\rm QCD}}{m_b} \right)^{3/2} 
   \,\,[\approx 0.5] \,.
\end{equation}
However, it is unclear whether the failure of power counting can be 
attributed to the form factors and decay constants alone.

Note that for the purposes of power counting we treated the charm quark 
as heavy, taking the heavy-quark limit for fixed $m_c/m_b$. This 
simplified the discussion, since we did not have to introduce $m_c$ as 
a separate scale. However, in reality charm is somewhat intermediate 
between a heavy and a light quark, since $m_c$ is not particularly large 
compared to $\Lambda_{\rm QCD}$. In this context it is worth noting that 
the first hard-scattering kernel in (\ref{fff}) cannot have 
$\Lambda_{\rm QCD}/m_c$ corrections, since there is a smooth transition 
to the case of two light mesons. The situation is different with the 
hard spectator interaction term, which we argued to be power suppressed 
for decays into a $D$ meson and a light meson. We shall come back to 
this in Sect.~\ref{bdpi2}, where we estimate the magnitude of this term 
for the $D\pi$ final state, relaxing the assumption that the $D$ meson 
is heavy.

\boldmath
\section{Phenomenology of $B\to D^{(*)} L$ decays}
\unboldmath
\label{bdpi}

The matrix elements we have computed in Sect.~\ref{menlo} provide
the theoretical basis for a model-independent calculation of the
class-I non-leptonic decay amplitudes for decays of the type
$B\to D^{(*)} L$, where $L$ is a light meson, to leading power
in $\Lambda_{\rm QCD}/m_b$ and at next-to-leading order in 
renormalization-group improved perturbation theory. In this section 
we discuss phenomenological applications of this formalism and 
confront our numerical results with experiment. We also provide
some numerical estimates of power-suppressed corrections to the 
factorization formula.

\subsection{Non-leptonic decay amplitudes}

The results for the class-I decay amplitudes for $B\to D^{(*)} L$  
are obtained by evaluating the (factorized) hadronic matrix elements 
of the transition operator ${\cal T}$ defined in (\ref{heffa1}). They 
are written in terms of products of CKM matrix elements, light-meson 
decay constants, $B\to D^{(*)}$ transition form factors, and the QCD 
parameters $a_1(D^{(*)} L)$. The decay constants can be determined 
experimentally using data on the weak leptonic decays 
$P^-\to l^-\bar\nu_l(\gamma)$, hadronic $\tau^-\to M^-\nu_\tau$ decays, 
and the electromagnetic decays $V^0\to e^+ e^-$. Following~\cite{NeSt97},
we use $f_\pi=131$\,MeV, $f_K=160$\,MeV, $f_\rho=210$\,MeV, 
$f_{K^*}=214$\,MeV, and $f_{a_1}=229$\,MeV. (Here $a_1$ is the 
pseudovector meson with mass $m_{a_1}\simeq 1230$\,MeV.)

The non-leptonic $\bar B_d\to D^{(*)+} L^-$ decay amplitudes for 
$L=\pi$, $\rho$ can be expressed as
\begin{eqnarray}\label{amplitudes}
   {\cal A}(\bar B_d\to D^+\pi^-)
   &=& i\frac{G_F}{\sqrt{2}}\,V^*_{ud}V_{cb}\,a_1(D\pi)\,
    f_\pi\,F_0(m^2_\pi)\,(m^2_B-m^2_D) \,, \nonumber\\
   {\cal A}(\bar B_d\to D^{*+}\pi^-)
   &=& - i\frac{G_F}{\sqrt{2}}\,V^*_{ud}V_{cb}\,a_1(D^*\pi)\,
    f_\pi A_0(m^2_\pi)\,2m_{D^*}\,\varepsilon^*\!\cdot p \,,
    \nonumber\\
   {\cal A}(\bar B_d\to D^{+}\rho^-)
   &=& - i\frac{G_F}{\sqrt{2}}\,V^*_{ud}V_{cb}\,a_1(D\rho)\,
    f_\rho\,F_+(m^2_\rho)\,2m_\rho\,\eta^*\!\cdot p \,,
\end{eqnarray}
where $p$ ($p'$) is the momentum of the $B$ (charm) meson, $\varepsilon$ 
and $\eta$ are polarization vectors, and the form factors $F_0$, $F_+$ 
and $A_0$ are defined in the usual way \cite{NeSt97}. The decay mode 
$\bar B_d\to D^{*+}\rho^-$ has a richer structure than the decays with 
at least one pseudoscalar in the final state. The most general 
Lorentz-invariant decomposition of the corresponding decay amplitude 
can be written as
\begin{equation}\label{abdsrho}
   {\cal A}(\bar B_d\to D^{*+}\rho^-) = i\frac{G_F}{\sqrt{2}}\,
   V^*_{ud} V_{cb}\,\varepsilon^{*\mu}\eta^{*\nu} \bigg(
   S_1\,g_{\mu\nu} - S_2\,q_\mu p'_\nu
   + iS_3\,\epsilon_{\mu\nu\alpha\beta}\,p'^\alpha q^\beta \bigg) \,,
\end{equation}
where the quantities $S_i$ can be expressed in terms of semi-leptonic
form factors. To leading power in $\Lambda_{\rm QCD}/m_b$, we obtain
\begin{eqnarray}
   S_1 &=& a_1(D^*\rho)\,m_\rho f_\rho\,(m_B+m_{D^*}) A_1(m_\rho^2)
    \,, \nonumber\\
   S_2 &=& a_1(D^*\rho)\,m_\rho f_\rho\,
    \frac{2 A_2(m_\rho^2)}{m_B+m_{D^*}} \,.
\end{eqnarray}
The contribution proportional to $S_3$ in (\ref{abdsrho}) is associated
with transversely polarized $\rho$ mesons and thus leads to 
power-suppressed effects, which we do not consider here. 

The various $B\to D^{(*)}$ form factors entering the expressions for 
the decay amplitudes can be determined by combining experimental data 
on semi-leptonic decays with theoretical relations derived using 
heavy-quark effective theory \cite{IW89,NeSt97}. Since we work to 
leading order in $\Lambda_{\rm QCD}/m_b$, it is consistent to set the 
light meson masses to zero and evaluate these form factors at $q^2=0$. 
In this case the kinematic relations
\begin{equation}\label{kinerela}
   F_0(0) = F_+(0) \,, \qquad
   (m_B+m_{D^*}) A_1(0) - (m_B-m_{D^*}) A_2(0) = 2 m_{D^*} A_0(0)
\end{equation}
allow us to express the two $\bar B_d\to D^+ L^-$ rates in terms of 
$F_+(0)$, and the two $\bar B_d\to D^{*+} L^-$ rates in terms of 
$A_0(0)$. Heavy-quark symmetry implies that these two form factors are 
equal to within a few percent \cite{review}. Below we adopt the common 
value $F_+(0)=A_0(0)=0.6$. All our predictions for decay rates will be 
proportional to the square of this number.

\subsection{Meson distribution amplitudes and predictions for $a_1$}

Let us now discuss in more detail the ingredients required for the
numerical analysis of the coefficients $a_1(D^{(*)} L)$. The Wilson 
coefficients $C_i$ in the effective weak Hamiltonian depend on the 
choice of the scale $\mu$ as well as on the value of the strong 
coupling $\alpha_s$, for which we take $\alpha_s(m_Z)=0.118$ and 
two-loop evolution down to a scale $\mu\sim m_b$. To study the 
residual scale dependence of the results, which remains because the 
perturbation series are truncated at next-to-leading order, we vary 
$\mu$ between $m_b/2$ and $2m_b$. The hard-scattering kernels depend
on the ratio of the heavy-quark masses, for which we take 
$z=m_c/m_b=0.30\pm 0.05$.

\begin{table}
\caption{\label{tab:wfnint}
Numerical values for the integrals $\int^1_0 du\,F(u,z)\,\Phi_L(u)$ 
(upper portion) and $\int^1_0 du\,F(u,-z)\,\Phi_L(u)$ (lower portion) 
obtained including the first two Gegenbauer moments.}
\vspace{0.2cm}
\begin{center}
\begin{tabular}{|c|c|c|c|}
\hline\hline
&&&\\[-0.35cm]
$z$ & Leading term & Coefficient of $\alpha_1^L$ &
 Coefficient of $\alpha_2^L$ \\
\hline
&&&\\[-0.35cm]
0.25 & $-8.41-9.51i$ & $5.92-12.19i$ & $-1.33+0.36i$ \\
0.30 & $-8.79-9.09i$ & $5.78-12.71i$ & $-1.19+0.58i$ \\
0.35 & $-9.13-8.59i$ & $5.60-13.21i$ & $-1.00+0.73i$ \\
\hline
&&&\\[-0.35cm]
0.25 & $-8.45-6.56i$ & $6.72-10.73i$ & $-0.38+0.93i$ \\
0.30 & $-8.37-5.99i$ & $6.83-11.49i$ & $-0.21+0.85i$ \\
0.35 & $-8.24-5.44i$ & $6.81-12.29i$ & $-0.08+0.75i$ \\
\hline\hline
\end{tabular}
\end{center}
\end{table}

Hadronic uncertainties enter the analysis also through the 
parameterizations used for the meson light-cone distribution amplitudes. 
It is convenient and conventional to expand the distribution amplitudes 
in Gegenbauer polynomials as
\begin{equation}\label{gpol}
   \Phi_L(u) = 6u(1-u) \left[ 1 + \sum_{n=1}^\infty 
   \alpha_n^L(\mu)\,C_n^{(3/2)}(2u-1) \right] \,,
\end{equation}
where $C_1^{(3/2)}(x)=3x$, $C_2^{(3/2)}(x)=\frac32(5x^2-1)$, etc. The 
Gegenbauer moments $\alpha_n^L(\mu)$ are multiplicatively renormalized. 
The scale dependence of these quantities would, however, enter the 
results for the coefficients only at order $\alpha_s^2$, which is 
beyond the accuracy of our calculation. We assume that the leading-twist 
distribution amplitudes are close to their asymptotic form and thus 
truncate the expansion at $n=2$. However, it would be straightforward 
to account for higher-order terms if desired. For the asymptotic form 
of the distribution amplitude, $\Phi_L(u)=6u(1-u)$, the integral in 
(\ref{a1dpi}) yields
\begin{eqnarray}\label{fintas}
   &&\int^1_0 du\,F(u,z)\,\Phi_L(u) 
    = 3\ln z^2 - 7 \nonumber\\
   &&\quad \mbox{}+ \Bigg[ \frac{6z(1-2z)}{(1-z)^2(1+z)^3}
    \left( \frac{\pi^2}{6} - \mbox{Li}_2(z^2) \right) 
    - \frac{3(2-3z+2z^2+z^3)}{(1-z)(1+z)^2} \ln(1-z^2) \nonumber\\
   &&\qquad\hspace*{0.3cm} \mbox{}+
    \frac{4-17z+20z^2+5z^3}{2(1-z)(1+z)^2} + \{ z\to 1/z\} \Bigg] \,,
\end{eqnarray}
and the corresponding result with the function $F(u,-z)$ is obtained by 
replacing $z\to -z$. More generally, a numerical integration with a 
distribution amplitude expanded in Gegenbauer polynomials yields the 
results collected in Table~\ref{tab:wfnint}. We observe that the first 
two Gegenbauer polynomials in the expansion of the light-cone 
distribution amplitudes give contributions of similar magnitude, 
whereas the second moment gives rise to much smaller effects. This 
tendency persists in higher orders. For our numerical discussion it 
is a safe approximation to truncate the expansion after the first 
non-trivial moment. The dependence of the results on the value of the 
quark mass ratio $z=m_c/m_b$ is mild and can be neglected for all 
practical purposes. We also note that the difference of the convolutions 
with the kernels for a pseudoscalar $D$ and vector $D^*$ meson are 
numerically very small. This observation is, however, specific to the 
case of $B\to D^{(*)} L$ decays and should not be generalized to other
decays.

{\tabcolsep=0.1cm
\begin{table}
\caption{\label{tab:a1dpi}
The QCD coefficients $a_1(D^{(*)} L)$ at next-to-leading order for 
three different values of the renormalization scale $\mu$. The 
leading-order values are shown for comparison.}
\vspace{0.2cm}
\begin{center}
\begin{tabular}{|c|ccc|}
\hline\hline
&&&\\[-0.35cm]
& $\mu=m_b/2$ & $\mu=m_b$ & $\mu=2 m_b$ \\
\hline
&&&\\[-0.35cm]
$a_1(D L)$ & $1.074+0.037i$ & $1.055+0.020i$ & $1.038+0.011i$ \\
 & $-(0.024-0.052i)\,\alpha_1^L$ & $-(0.013-0.028i)\,\alpha_1^L$ &
 $-(0.007-0.015i)\,\alpha_1^L$ \\
&&&\\[-0.35cm]
$a_1(D^* L)$ & $1.072+0.024i$ & $1.054+0.013i$ & $1.037+0.007i$ \\
 & $-(0.028-0.047i)\,\alpha_1^L$ & $-(0.015-0.025i)\,\alpha_1^L$ &
 $-(0.008-0.014i)\,\alpha_1^L$ \\
&&&\\[-0.35cm]
$a^{\rm LO}_1$ & $1.049$ & $1.025$ & $1.011$ \\
\hline\hline
\end{tabular}
\end{center}
\end{table}}

Next we evaluate the complete results for the parameters $a_1$ at
next-to-leading order, and to leading power in $\Lambda_{\rm QCD}/m_b$. 
We set $z=m_c/m_b=0.3$. Varying $z$ between 0.25 and 0.35 would change
the results by less than 0.5\%. The results are shown in 
Table~\ref{tab:a1dpi}. The contributions proportional to the
second Gegenbauer moment $\alpha_2^L$ have coefficients of order 0.2\% 
or less and can safely be neglected.
The contributions associated with $\alpha_1^L$ are present only for the
strange mesons $K$ and $K^*$, but not for $\pi$ and $\rho$.  Moreover,
the imaginary parts of the coefficients contribute to their modulus
only at order $\alpha_s^2$, which is beyond the accuracy of our 
analysis. To summarize, we thus obtain
\begin{eqnarray}\label{a1mods}
   |a_1(D L)| &=& 1.055_{-0.017}^{+0.019}
    - (0.013_{-0.006}^{+0.011})\alpha_1^L \,, \nonumber\\
   |a_1(D^* L)| &=& 1.054_{-0.017}^{+0.018}
    - (0.015_{-0.007}^{+0.013})\alpha_1^L \,, 
\end{eqnarray}
where the quoted errors reflect the perturbative uncertainty due to 
the scale ambiguity (and the negligible dependence on the value of the 
ratio of quark masses and higher Gegenbauer moments), but not the 
effects of power-sup\-pressed corrections. These will be estimated 
later. It is evident that within theoretical uncertainties there is no 
significant difference between the two $a_1$ parameters, and there is 
only a very small sensitivity to the differences between strange and 
non-strange mesons (assuming that $|\alpha_1^{K^{(*)}}|<1$). In our 
numerical analysis below we thus take $|a_1|=1.05$ for all decay modes.

\subsection{Tests of factorization}

The main lesson from the previous discussion is that corrections to
naive factorization in the class-I decays $\bar B_d\to D^{(*)+} L^-$ 
are very small. The reason is that these effects are governed by a 
small Wilson coefficient and, moreover, are colour suppressed by a 
factor $1/N_c^2$. For these decays, the most important implications of 
the QCD factorization formula are to restore the renormalization-group 
invariance of the theoretical predictions, and to provide a theoretical 
justification for why naive factorization works so well. On the other 
hand, given the theoretical uncertainties arising, e.g., from unknown 
power-suppressed corrections, there is little hope to confront the 
extremely small predictions for non-universal (process-dependent) 
``non-factorizable'' corrections with experimental data. Rather, what 
we may do is ask whether data supports the prediction of a 
quasi-universal parameter $|a_1|\simeq 1.05$ in these decays. If this 
is indeed the case, it would support the usefulness of the heavy-quark 
limit in analyzing non-leptonic decay amplitudes. If, on the other hand, 
we were to find large non-universal effects, this would point towards 
the existence of sizeable power corrections to our predictions. We 
will see that within present experimental errors the data are in good
agreement with our prediction of a quasi universal $a_1$ parameter. 
However, a reduction of the experimental uncertainties to the percent 
level would be very desirable for obtaining a more conclusive picture.

We start by considering ratios of non-leptonic decay rates that are 
related to each other by the
replacement of a pseudoscalar meson by a vector meson. In the comparison
of $B\to D\pi$ and $B\to D^*\pi$ decays one is sensitive to the 
difference of the values of the two $a_1$ parameters in (\ref{a1mods})
evaluated for $\alpha_1^L=0$. This difference is at most few times 
$10^{-3}$. Likewise, in the comparison of $B\to D\pi$ and $B\to D\rho$ 
decays one is sensitive to the difference in the light-cone distribution 
amplitudes of the pion and the $\rho$ meson, which start at the second 
Gegenbauer moment $\alpha_2^L$. These effects are suppressed even more 
strongly. From the explicit expressions for the decay amplitudes in 
(\ref{amplitudes}) it follows that
\begin{eqnarray}
   \frac{\Gamma(\bar B_d\to D^+\pi^-)}{\Gamma(\bar B_d\to D^{*+}\pi^-)}
   &=& \frac{(m_B^2-m_D^2)^2|\vec q\,|_{D\pi}}
            {4m_B^2|\vec q\,|_{D^*\pi}^3}
    \left( \frac{F_0(m_\pi^2)}{A_0(m_\pi^2)} \right)^2
    \left| \frac{a_1(D\pi)}{a_1(D^*\pi)} \right|^2 , \nonumber\\
   \frac{\Gamma(\bar B_d\to D^+\rho^-)}{\Gamma(\bar B_d\to D^+\pi^-)}
   &=& \frac{4m_B^2|\vec q\,|_{D\rho}^3}
            {(m_B^2-m_D^2)^2|\vec q\,|_{D\pi}}\,
    \frac{f_\rho^2}{f_\pi^2}\,
    \left( \frac{F_+(m_\rho^2)}{F_0(m_\pi^2)} \right)^2
    \left| \frac{a_1(D\rho)}{a_1(D\pi)} \right|^2 .~~
\end{eqnarray}
Using the experimental values for the branching ratios reported by the
CLEO Collaboration \cite{CLEO9701} we find (taking into account a 
correlation between some systematic errors in the second case)
\begin{eqnarray}
   \left| \frac{a_1(D\pi)}{a_1(D^*\pi)} \right|\,
   \frac{F_0(m_\pi^2)}{A_0(m_\pi^2)}
   &=& 1.00\pm 0.11 \,, \nonumber\\
   \left| \frac{a_1(D\rho)}{a_1(D\pi)} \right|\,
   \frac{F_+(m_\rho^2)}{F_0(m_\pi^2)}
   &=& 1.16\pm 0.11 \,.
\end{eqnarray}
Within errors, there is no evidence for any deviations from naive 
factorization.

Our next-to-leading order results for the quantities $a_1(D^{(*)} L)$ 
allow us to make theoretical predictions which are not restricted to 
ratios of hadronic decay rates. A particularly clean test of these 
predictions, which is essentially free of hadronic uncertainties, is 
obtained by relating the $\bar B_d\to D^{(*)+} L^-$ decay rates to the 
differential semi-leptonic $\bar B_d\to D^{(*)+}\,l^-\nu$ decay rate 
evaluated at $q^2=m_L^2$. In this way the parameters $|a_1|$ can be 
measured directly \cite{Bj89}. One obtains
\begin{equation}\label{tfrpi}
   R_L^{(*)} = \frac{\Gamma(\bar B_d\to D^{(*)+} L^-)}
    {d\Gamma(\bar B_d\to D^{(*)+} l^-\bar\nu)/dq^2\big|_{q^2=m^2_L}}
   = 6\pi^2 |V_{ud}|^2 f^2_L\,|a_1(D^{(*)} L)|^2\,X^{(*)}_L \,,
\end{equation}
where $X_\rho=X_\rho^*=1$ for a vector meson (because the production
of the lepton pair via a $V-A$ current in semi-leptonic decays is
kinematically equivalent to that of a vector meson with momentum $q$), 
whereas $X_\pi$ and $X_\pi^*$ deviate from 1 only by (calculable) terms 
of order $m_\pi^2/m_B^2$, which numerically are below the 1\% level
\cite{NeSt97}. We emphasize that with our results for $a_1$ given in 
(\ref{a1dpi}) the above relation becomes a prediction based on first 
principles of QCD. This is to be contrasted with the usual 
interpretation of this formula, where $a_1$ plays the role of a 
phenomenological parameter that is fitted from data.

The most accurate tests of factorization employ the 
class-I processes $\bar B_d\to D^{*+}L^-$, because the differential 
semi-leptonic decay rate in $B\to D^*$ transitions has been measured as 
a function of $q^2$ with good accuracy. The results of such an analysis, 
performed using CLEO data, have been reported in \cite{Rodr97}. One 
finds
\begin{eqnarray}\label{a1exp}
   R_\pi^* = (1.13\pm 0.15)\,\mbox{GeV}^2
   \quad &\Rightarrow& \quad
    |a_1(D^*\pi)| = 1.08 \pm 0.07 \,, \nonumber\\
   R_\rho^* = (2.94\pm 0.54)\,\mbox{GeV}^2
   \quad &\Rightarrow& \quad
    |a_1(D^*\rho)| = 1.09\pm 0.10 \,, \nonumber\\
   R_{a_1}^* = (3.45\pm 0.69)\,\mbox{GeV}^2
   \quad &\Rightarrow& \quad
    |a_1(D^* a_1)| = 1.08\pm 0.11 \,.
\end{eqnarray}
This is consistent with our theoretical result in (\ref{a1dpi}). In
particular, the data show no evidence for large power corrections to 
our predictions obtained at leading order in $\Lambda_{\rm QCD}/m_b$. 
However, a further improvement in the experimental accuracy would be 
desirable in order to become sensitive to process-dependent, 
non-factorizable effects.

\subsection{Predictions for class-I decay amplitudes}

We now consider a larger set of class-I decays of the form 
$\bar B_d\to D^{(*)+} L^-$, all of which are governed by the transition 
operator (\ref{heffa1}). In Table~\ref{tab:10decays} we compare the QCD 
factorization predictions with experimental data. As previously we work 
in the heavy-quark limit, i.e.\ our predictions are model independent 
up to corrections suppressed by at least one power of 
$\Lambda_{\rm QCD}/m_b$. The results show good agreement with experiment 
within errors, which are still rather large. (Note that we have 
not attempted to adjust the semi-leptonic form factors $F_+(0)$ and 
$A_0(0)$ so as to obtain a best fit to the data.)

\begin{table}
\caption{\label{tab:10decays}
Model-independent predictions for the branching ratios (in units of
$10^{-3}$) of class-I, non-leptonic $\bar B_d\to D^{(*)+} L^-$ decays 
in the heavy-quark limit. All predictions are in units of 
$(|a_1|/1.05)^2$. The last two columns show the experimental results 
reported by the CLEO Collaboration \protect\cite{CLEO9701}, and by the 
Particle Data Group \protect\cite{PDG}.}
\vspace{0.2cm}
\begin{center}
\begin{tabular}{|l|c|cc|}
\hline\hline
&&&\\[-0.35cm]
Decay mode & Theory (HQL) & CLEO data & PDG98~ \\
\hline
&&&\\[-0.35cm]
$\bar B_d\to D^+\pi^-$   & 3.27 & $2.50\pm 0.40$ & $3.0\pm 0.4$ \\
$\bar B_d\to D^+ K^-$    & 0.25  & --- & --- \\
$\bar B_d\to D^+\rho^-$  & 7.64  & $7.89\pm 1.39$ & $7.9\pm 1.4$ \\
$\bar B_d\to D^+ K^{*-}$ & 0.39  & --- & --- \\
$\bar B_d\to D^+ a_1^-$  & 7.76  & $8.34\pm 1.66$ & $6.0\pm 3.3$ \\
 & $\times[F_+(0)/0.6]^2$ & & \\
\hline
&&&\\[-0.35cm]
$\bar B_d\to D^{*+}\pi^-$   & 3.05  & $2.34\pm 0.32$ & $2.8\pm 0.2$ \\
$\bar B_d\to D^{*+} K^-$    & 0.22  & --- & --- \\
$\bar B_d\to D^{*+}\rho^-$  & 7.59  & $7.34\pm 1.00$ & $6.7\pm 3.3$ \\
$\bar B_d\to D^{*+} K^{*-}$ & 0.40 & --- & --- \\
$\bar B_d\to D^{*+} a_1^-$  & 8.53 & $11.57\pm 2.02$ & $13.0\pm 2.7$ \\
 & $\times[A_0(0)/0.6]^2$ & & \\
\hline\hline
\end{tabular}
\end{center}
\end{table}

We take the observation that the experimental data on
class-I decays into heavy-light final states show good agreement with 
our predictions obtained in the heavy-quark limit as evidence 
that in these decays there are no unexpectedly large power corrections.
We will now address the important question of the size of power 
corrections theoretically. To this end we provide rough estimates of
two sources of power-suppressed effects: weak annihilation and spectator
interactions. We stress that, at present, a complete account of power 
corrections to the heavy-quark limit cannot be performed in a systematic 
way, since these effects are not dominated by hard gluon exchange. In 
other words, factorization breaks down beyond leading power, and there
are other sources of power corrections, such as contributions from 
higher Fock states, which we will not address here. We believe that the 
estimates presented below are nevertheless instructive. 

To obtain an estimate of power corrections we adopt the following, 
heuristic procedure. We treat the charm quark as {\em light\/} compared 
to the large scale provided by the mass of the decaying $b$ quark 
($m_c\ll m_b$, and $m_c$ {\em fixed\/} as $m_b\to\infty$) and use a 
light-cone projection 
similar to that of the pion also for the $D$ meson. In addition, we 
assume that $m_c$ is still large compared to $\Lambda_{\rm QCD}$. We 
implement this by using a highly asymmetric $D$-meson wave function, 
which is strongly peaked at a light-quark momentum fraction of order 
$\Lambda_{\rm QCD}/m_D$. This guarantees correct power counting for the 
heavy-light final states we are interested in. As discussed in 
Sect.~\ref{subsec:annihilation}, there are four annihilation diagrams 
with a single gluon exchange (see Fig.~\ref{fig9} (a)--(d)). The first 
two diagrams are ``factorizable'' and their contributions vanish 
because of current conservation in the limit $m_c\to 0$. For non-zero 
$m_c$ they therefore carry an additional suppression factor 
$m^2_D/m^2_B\approx 0.1$. Moreover, their contributions to the decay 
amplitude are suppressed by small Wilson coefficients. Diagrams (a) and 
(b) can therefore safely be neglected. From the non-factorizable 
diagrams (c) and (d) in Fig.~\ref{fig9}, the one with the gluon 
attached to the $b$ quark turns out to be strongly suppressed 
numerically, giving a contribution of less than $1\%$ of the leading 
class-I amplitude. We are thus left with diagram (d), in which the gluon 
couples to the light quark in the $B$ meson. This mechanism gives the 
dominant annihilation contribution. (Note that by deforming the light 
spectator-quark line one can redraw this diagram in such a way that it 
can be interpreted as a final-state rescattering process.) 

Adopting a common notation, we parameterize the annihilation contribution 
to the $\bar B_d\to D^+\pi^-$ decay amplitude in terms of a 
(power-suppressed) amplitude $A$ such that 
${\cal A}(\bar B_d\to D^+\pi^-)=T+A$, 
where $T$ is the ``tree topology'', which contains the dominant 
factorizable contribution. A straightforward calculation using the 
approximations discussed above shows that the contribution of diagram 
(d) is (to leading order) independent of the momentum fraction $\xi$ of 
the light quark inside the $B$ meson:
\begin{equation}\label{wa9d}
   A \simeq f_\pi f_D f_B \int du\,\frac{\Phi_\pi(u)}{u}
   \int dv\,\frac{\Phi_D(v)}{\bar v^2} 
   \simeq 3 f_\pi f_D f_B\,\int dv\,\frac{\Phi_D(v)}{\bar v^2} \,.
\end{equation}
The $B$-meson wave function simply integrates to $f_B$, and the 
integral over the pion distribution amplitude can be performed using
the asymptotic form of the wave function. We take $\Phi_D(v)$ in the 
form of (\ref{gpol}) with the coefficients $\alpha^D_1=0.8$ and 
$\alpha^D_2=0.4$ ($\alpha^D_i=0$, $i>2$). With this ansatz $\Phi_D(v)$ 
is strongly peaked at $\bar v\sim\Lambda_{\rm QCD}/m_D$. The integral 
over $\Phi_D(v)$ in (\ref{wa9d}) is divergent at $v=1$, and we regulate 
it by introducing a cut-off such that $v\le 1-\Lambda/m_B$ with 
$\Lambda\approx 0.3$\,GeV. Then $\int dv\,\Phi_D(v)/\bar v^2\approx 34$. 
Evidently, the proper value of $\Lambda$ is largely unknown, and our 
estimate will be correspondingly uncertain. Nevertheless, this exercise 
will give us an idea of the magnitude of the effect. For the ratio of 
the annihilation amplitude to the leading, factorizable contribution we 
obtain
\begin{equation}\label{ws1}
   \frac{A}{T} \simeq \frac{2\pi\alpha_s}{3}\,
   \frac{C_+ + C_-}{2C_+ + C_-}\,
   \frac{f_D f_B}{F_0(0)\,m_B^2}\int dv\,\frac{\Phi_D(v)}{\bar v^2}
   \approx 0.04 \,.
\end{equation}
We have evaluated the Wilson coefficients at $\mu=m_b$ and used 
$f_D=0.2$\,GeV, $f_B=0.18$\,GeV, $F_0(0)=0.6$, and $\alpha_s=0.4$.
This value of the strong coupling constant reflects that the typical 
virtuality of the gluon propagator in the annihilation graph is of 
order $\Lambda_{\rm QCD} m_B$. We conclude that the annihilation 
contribution is a correction of a few percent, which is what one would
expect for a generic power correction to the heavy-quark limit. Taking 
into account that $f_B\sim\Lambda_{\rm QCD}(\Lambda_{\rm QCD}/m_B)^{1/2}$,
$F_0(0)\sim(\Lambda_{\rm QCD}/m_B)^{3/2}$ and $f_D\sim\Lambda_{\rm QCD}$,
we observe that in the heavy-quark limit the ratio $A/T$ indeed scales 
as $\Lambda_{\rm QCD}/m_b$, exhibiting the expected linear power 
suppression. (Recall that we consider the $D$ meson as a light meson 
for this heuristic analysis of power corrections.)

Using the same approach, we may also derive a numerical estimate for 
the non-factorizable spectator interaction in $\bar B_d\to D^+\pi^-$ 
decays, discussed in Sect.~\ref{subsec:hardspec}. We find
\begin{equation}\label{estnfs}
   \frac{T_{\rm spec}}{T_{\rm lead}}
   \simeq \frac{2\pi\alpha_s}{3}\,
   \frac{C_+ - C_-}{2C_+ + C_-}\,
   \frac{f_D f_B}{F_0(0)\,m_B^2}\frac{m_B}{\lambda_B}
   \int dv\,\frac{\Phi_D(v)}{\bar v} \approx -0.03 \,,
\end{equation}
where the hadronic parameter $\lambda_B=O(\Lambda_{\rm QCD})$ is 
defined as $\int_0^1(d\xi/\xi)\,\Phi_B(\xi)\equiv m_B/\lambda_B$.
For the numerical estimate we have assumed that 
$\lambda_B\approx 0.3$\,GeV. With the same model for $\Phi_D(v)$ as 
above we have $\int dv\,\Phi_D(v)/\bar v\approx 6.6$, where the integral 
is now convergent. The result (\ref{estnfs}) exhibits again the expected 
power suppression in the heavy-quark limit, and the numerical size of 
the effect is at the few percent level.

We conclude from this discussion that the typical size of power
corrections to the heavy-quark limit in class-I decays of $B$ mesons
into heavy-light final states is at the level of 10\% or less, and
thus our prediction for the near universality of the parameters $a_1$
governing these decay modes appears robust. 

\subsection{Remarks on class-II and class-III decay amplitudes}
\label{bdpi2}

In the class-I decays $\bar B_d\to D^{(*)+} L^-$, the flavour quantum 
numbers of the final-state mesons ensure that only the light meson $L$ 
can be produced by the $(\bar d u)$ current contained in the operators 
of the effective weak Hamiltonian in (\ref{heff18}). The QCD 
factorization formula then predicts that the corresponding decay 
amplitudes are factorizable in the heavy-quark limit. The formula also 
predicts that other topologies, in which the heavy charm meson would be 
created by a $(\bar c u)$ current, are power suppressed. To study these 
topologies we now consider decays with a neutral charm meson in the 
final state. In the class-II decays $\bar B_d\to D^{(*)0} L^0$ the only 
possible topology is to have the charm meson as the emission particle, 
whereas for the class-III decays $B^-\to D^{(*)0} L^-$ both final-state 
mesons can be the emission particle. The factorization formula predicts 
that in the heavy-quark limit class-II decay amplitudes are power 
suppressed with respect to the corresponding class-I amplitudes, whereas 
class-III amplitudes should be equal to the corresponding class-I 
amplitudes up to power corrections. 

It is convenient to introduce two common parameterizations of the 
decay amplitudes, one in terms of isospin amplitudes $A_{1/2}$ and 
$A_{3/2}$ referring to the isospin of the final-state particles, and 
one in terms of flavour topologies ($T$ for ``tree topology'', $C$ for
``colour suppressed tree topology'', and $A$ for ``annihilation 
topology''). Taking the decays $B\to D\pi$ as an example, we have
\begin{eqnarray}\label{asw}
   {\cal A}(\bar B_d\to D^+\pi^-)
   &=& \sqrt{\frac13} A_{3/2} + \sqrt{\frac23} A_{1/2}
    = T + A \,, \nonumber\\[0.1cm]
   \sqrt2\,{\cal A}(\bar B_d\to D^0\pi^0) 
   &=& \sqrt{\frac43} A_{3/2} - \sqrt{\frac23} A_{1/2}
    = C - A \,, \nonumber\\[0.3cm]
   {\cal A}(B^-\to D^0\pi^-) &=& \sqrt3 A_{3/2} = T + C \,.
\end{eqnarray}
A similar decomposition holds for the other $B\to D^{(*)} L$ decay 
modes. Isospin symmetry of the strong interactions implies that the 
class-III amplitude is a linear combination of the class-I and class-II 
amplitudes. In other words, there are only two independent amplitudes, 
which can be taken to be $A_{1/2}$ and $A_{3/2}$, or $(T+A)$ and 
$(C-A)$. These amplitudes are complex due to strong-interaction phases 
from final-state interactions. Only the relative phase of the two 
independent amplitudes is an observable. We define $\delta$ to be the 
relative phase of $A_{1/2}$ and $A_{3/2}$, and $\delta_{TC}$ the 
relative phase of $(T+A)$ and $(C-A)$. The QCD factorization formula 
implies that
\begin{eqnarray}\label{delTC}
   \frac{A_{1/2}}{\sqrt2\,A_{3/2}}
   &=& 1 + O(\Lambda_{\rm QCD}/m_b) \,, \qquad
    \delta = O(\Lambda_{\rm QCD}/m_b) \,, \nonumber\\
   \frac{C-A}{T+A}
   &=& O(\Lambda_{\rm QCD}/m_b) \,, \hspace{0.9cm}
    \delta_{TC} = O(1) \,.
\end{eqnarray}
In the remainder of this section, we will explore to what extent these 
predictions are supported by data.

{\tabcolsep=0.18cm
\begin{table}
\caption{\label{tab:brbdpi}
CLEO data \protect\cite{CLEO9701,CLEOsupp} on the branching ratios for 
the decays $B\to D^{(*)} L$ in units of $10^{-3}$. Upper limits are at 
90\% confidence level. See text for the definition of the quantities 
$\delta$ and ${\cal R}$.}
\vspace{0.2cm}
\begin{center}
\begin{tabular}{|c|cccc|}
\hline\hline
&&&&\\[-0.35cm]
 & $B\to D\pi$ & $B\to D\rho$ & $B\to D^*\pi$ & $B\to D^*\rho$ \\
\hline
&&&&\\[-0.35cm]
Class-I~~ ($D^{(*)+} L^-$) & $2.50\pm 0.40$ & $ 7.89\pm 1.39$
 & $2.34\pm 0.32$ & $7.34\pm 1.00$ \\
Class-II~~ ($D^{(*)0} L^0$) & $<0.12$ & $<0.39$ & $<0.44$ & $<0.56$ \\
Class-III ($D^{(*)0} L^-$) & $4.73\pm 0.44$ & $9.20\pm 1.11$
 & $3.92\pm 0.63$ & $12.77\pm 1.94$ \\
\hline
&&&&\\[-0.35cm]
$\delta$ & $<22^\circ$ & $<30^\circ$ & $<57^\circ$ & $<31^\circ$ \\
${\cal R}$ & $1.34\pm 0.13$ & $1.05\pm 0.12$ & $1.26\pm 0.14$ 
 & $1.28\pm 0.13$ \\
\hline\hline
\end{tabular}
\end{center}
\end{table}}

In Table~\ref{tab:brbdpi} we show the experimental results for the
various $B\to D^{(*)} L$ branching ratios reported by the CLEO
Collaboration \cite{CLEO9701,CLEOsupp}. We first note that no evidence
has been seen for any of the class-II decays, in accordance with our 
prediction that these decays are suppressed with respect to the class-I
modes. Below we will investigate in more detail how this suppression is 
realized. The fourth line in the table shows upper limits on the 
strong-interaction phase difference $\delta$ between the two isospin 
amplitudes, which follow from the relation~\cite{NeSt97}
\begin{equation}
   \sin^2\!\delta < \frac92\,\frac{\tau(B^-)}{\tau(B_d)}\,
   \frac{\mbox{Br}(\bar B_d\to D^0\pi^0)}
        {\mbox{Br}(B^-\to D^0\pi^-)} \,.
\end{equation}
The strongest bound arises in the decays $B\to D\pi$, where the 
strong-interaction phase is bound to be less than $22^\circ$. This 
confirms our prediction that the phase $\delta$ is suppressed in the 
heavy-quark limit.

Let us now study the suppression of the class-II amplitudes in more 
detail. We have already mentioned in Sect.~\ref{subsec:diffcharm} that 
the observed smallness of these amplitudes is more a reflection of 
colour suppression than power suppression. This is already apparent in 
the naive factorization approximation, because the appropriate ratios 
of meson decay constants and semi-leptonic form factors exhibit large 
deviations from their expected scaling laws in the heavy-quark limit, 
as shown in (\ref{scalingviolations}). Indeed, it is obvious from 
Table~\ref{tab:brbdpi} that there are significant differences between
the class-I and class-III amplitudes, indicating that some 
power-suppressed contributions are not negligible. In the last row of 
the table we show the experimental values of the quantity
\begin{equation}
   {\cal R} = \left|  
   \frac{{\cal A}(B^-\to D^{(*)0} L^-)}
          {{\cal A}(\bar B_d\to D^{(*)+} L^-)} \right|
   = \sqrt{ \frac{\tau(B_d)}{\tau(B^-)}\,
    \frac{\mbox{Br}(B^-\to D^{(*)0} L^-)}
         {\mbox{Br}(\bar B_d\to D^{(*)+} L^-)}} \,,
\end{equation}
which parameterizes the magnitude of power-suppressed effects at the
level of the decay amplitudes. If we ignore the decays $B\to D^*\rho$ 
with two vector mesons in the final state, which are more complicated 
because of the presence of different helicity amplitudes, then the 
ratio ${\cal R}$ is given by
\begin{equation}\label{Rdef}
   {\cal R} = \left| 1 + \frac{C-A}{T+A} \right|
   = \left| 1 + x\,\frac{a_2}{a_1} \right| \,,
\end{equation}
where $a_1$ are the QCD parameters entering the transition operator in 
(\ref{heffa1}), and
\begin{equation}
   a_2 = \frac{N_c+1}{2 N_c}\,C_+ - \frac{N_c-1}{2 N_c}\,C_-
   + \mbox{``non-factorizable corrections''} 
\end{equation}
are the corresponding parameters describing the deviations from naive 
factorization in the class-II decays \cite{NeSt97}. All quantities in 
(\ref{Rdef}) depend on the nature of the final-state mesons. In 
particular, the parameters
\begin{eqnarray}
   x(D\pi)
   &=& \frac{(m_B^2-m_\pi^2)\,f_D\,F_0^{B\to\pi}(m_D^2)}
            {(m_B^2-m_D^2)\,f_\pi\,F_0^{B\to D}(m_\pi^2)} 
    \approx 0.9 \,, \nonumber\\
   x(D\rho)
   &=& \frac{f_D\,A_0^{B\to\rho}(m_D^2)}
            {f_\rho\,F_+^{B\to D}(m_\rho^2)} 
    \approx 0.5 \,, \nonumber\\
   x(D^*\pi)
   &=& \frac{f_{D^*} F_+^{B\to\pi}(m_{D^*}^2)}
            {f_\pi\,A_0^{B\to D^*}(m_\pi^2)} 
    \approx 0.9 \,,
\end{eqnarray}
account for the ratios of decay constants and form factors entering
in the naive factorization approximation. For the numerical estimates 
we have assumed that the ratios of heavy-to-light over heavy-to-heavy 
form factors are approximately equal to 0.5, and we have taken 
$f_D=0.2$\,GeV and $f_{D^*}=0.23$\,GeV for the charm meson decay 
constants. Note that in (\ref{Rdef}) it is the quantities $x$ that are 
formally power suppressed $\sim(\Lambda_{\rm QCD}/m_b)^2$ in the 
heavy-quark limit, {\em not\/} the ratios $a_2/a_1$. For the final 
states containing a pion the power suppression is clearly not operative, 
mainly due to the fact that the pion decay constant $f_\pi$ is much 
smaller than the quantity $(f_D\sqrt{m_D})^{2/3}\approx 0.42$\,GeV. To 
reproduce the experimental values of the ratios ${\cal R}$ shown in 
Table~\ref{tab:brbdpi} requires values of $a_2/a_1$ of order 0.1--0.4 
(with large uncertainties), which is consistent with the fact that these 
ratios are of order $1/N_c$ in the large-$N_c$ limit, i.e.\ they are 
colour suppressed. 

The QCD factorization formula (\ref{fff}) allows us to compute the
coefficients $a_1$ in the heavy-quark limit, but it does {\em not\/}
allow us to compute the corresponding parameters $a_2$ in class-II 
decays. The reason is that in class-II decays the emission particle is
a heavy charm meson, and hence the mechanism of colour transparency,
which was essential for the proof of factorization, is not operative. 
For a rough estimate of $a_2$ in $B\to\pi D$ decays we consider 
as previously the limit in which the charm meson is treated as a light 
meson ($m_c\ll m_b$), however with a highly asymmetric distribution 
amplitude. In this limit we can adapt our results for the class-II 
amplitude in $B\to\pi\pi$ decays derived in \cite{BBNS99}, with the 
only modification that the hard-scattering kernels must be generalized 
to the case where the leading-twist light-cone distribution amplitude
of the emission meson is not symmetric. We find that
\begin{eqnarray}
   a_2 &\simeq& \frac{N_c+1}{2 N_c}\,\bar C_+(\mu)
    - \frac{N_c-1}{2 N_c}\,\bar C_-(\mu) \nonumber\\
   &&\mbox{}+ \frac{\alpha_s}{4\pi}\,\frac{C_F}{2 N_c}\,
    [\bar C_+(\mu)+\bar C_-(\mu)] \left( -6\ln\frac{\mu^2}{m_b^2}
    + f_I + f_{II} \right) ,
\end{eqnarray}
where
\begin{eqnarray}
   f_I &=& \int_0^1 dv\,\Phi_D(v) \left[
    \ln^2\!\bar v + \ln\bar v + \frac{\pi^2}{3} - 6
    + i\pi (2\ln\bar v - 3) + O(\bar v) \right] \,, \nonumber\\
   f_{II} &=& \frac{12\pi^2}{N_c}\,
    \frac{f_\pi f_B}{F_0^{B\to\pi}(m_D^2)\,m_B^2}\,
    \frac{m_B}{\lambda_B}\int dv \frac{\Phi_D(v)}{\bar v} \,.
\end{eqnarray}
The contribution from $f_{II}$ describes the hard, non-factorizable 
spectator interaction. Note that this term involves 
$\int dv\,\Phi_D(v)/\bar v$, which can be sizeable but remains 
constant in the heavy-quark limit implied here ($m_b\to\infty$ with  
$m_c$ constant). Using the same numerical inputs as previously, we find 
that $f_{II}\approx 13$ and $f_I\approx-1-19i$. In writing the 
hard-scattering kernel for $f_I$ we have only kept the leading terms in 
$\bar v$, which is justified because of the strongly asymmetric shape 
of $\Phi_D(v)$. Note the large imaginary part arising from the 
``non-factorizable'' vertex corrections with a gluon exchange between 
the final-state quarks. Combining all contributions, and taking 
$\mu=m_b$ for the renormalization scale, we find
\begin{equation}\label{a2est}
   a_2 \approx 0.25\,e^{-i 41^\circ} \,,
\end{equation}
which is significantly larger in magnitude than the leading-order 
result $a_2^{\rm LO}\approx 0.12$ corresponding to naive factorization. 
We hasten to add that our estimate (\ref{a2est}) should not be taken 
too seriously, since it is most likely not a good approximation to treat 
the charm meson as a light meson. Nevertheless, it is remarkable that in 
this idealized limit one obtains indeed a very significant correction to 
naive factorization, which gives the right order of magnitude for the 
modulus of $a_2$ and, at the same time, a large strong-interaction 
phase. For completeness, we note that the value for $a_2$ in 
(\ref{a2est}) would imply a strong-interaction phase difference 
$\delta\approx 10^\circ$ between the two isospin amplitudes $A_{1/2}$ 
and $A_{3/2}$ in $B\to D\pi$ decays, and hence is not in conflict with 
the experimental upper bound on the phase $\delta$ given in 
Table~\ref{tab:brbdpi}. The phase $\delta_{TC}$, on the other hand, is 
to leading order simply given by the phase of $a_2$ and is indeed large, 
in accordance with (\ref{delTC}).

\section{Conclusion}
\label{conclusion}

With the recent commissioning of the $B$ factories and the planned 
emphasis on heavy-flavour physics in future collider experiments, the 
role of $B$ decays in providing fundamental tests of the Standard Model 
and potential signatures of new physics will continue to grow. In many 
cases the principal source of systematic uncertainty is a theoretical 
one, namely our inability to quantify the non-perturbative QCD effects 
present in these decays. This is true, in particular, for almost all 
measurements of CP violation at the $B$ factories. 

In these lectures, I have reviewed a rigorous framework for the 
evaluation of strong-interaction effects for a large class of exclusive, 
two-body non-leptonic decays of $B$ mesons. The main result is contained 
in the factorization formula (\ref{fff}), which expresses the amplitudes 
for these decays in terms of experimentally measurable semi-leptonic 
form factors, light-cone distribution amplitudes, and hard-scattering 
functions that are calculable in perturbative QCD. For the first time, 
therefore, we have a well founded field-theoretic basis for 
phenomenological studies of exclusive hadronic $B$ decays, and a 
formal justification for the ideas of factorization. For simplicity, I 
have mainly focused on $B\to D\pi$ decays here. A detailed discussion 
of $B$ decays into two light mesons will be presented in a 
forthcoming paper \cite{BBNSfuture}.

We hope that the factorization formula (\ref{fff}) will form the basis
for future studies of non-leptonic two-body decays of $B$ mesons. 
Before, however, a fair amount of conceptual work remains to be 
completed. In particular, it will be important to investigate better the 
limitations on the numerical precision of the factorization formula, 
which is valid in the formal heavy-quark limit. We have discussed some 
preliminary estimates of power-suppressed effects in the present work, 
but a more complete analysis would be desirable. In particular, for 
rare $B$ decays into two light mesons it will be important to understand 
the role of chirally-enhanced power corrections and weak annihilation
contributions~\cite{BBNSfuture,Lietal}. For these decays, there 
are also still large uncertainties associated with the description of 
the hard spectator interactions. 

Theoretical investigations along these lines should be pursued with 
vigor. We are confident that, ultimately, this research will 
result in a {\em theory\/} of non-leptonic $B$ decays, which should be 
as useful for this area of heavy-flavour physics as the large-$m_b$ 
limit and heavy-quark effective theory were for the phenomenology of 
semi-leptonic decays.

\section*{Acknowledgements}

I would like to thank the organizers of the TASI Institute for the
invitation to present these lecture, for their hospitality, and for 
providing a stimulating atmosphere during the school. I am grateful to 
the students for attending the lectures and contributing with questions 
and discussions. Among many pleasant experiences during my stay in 
Boulder, I will remember a successful climb of Longs Peak, which helped
me to recover from the course. Finally, I am indebted to my 
collaborators Martin Beneke, Gerhard Buchalla and Chris Sachrajda, who
deserve much credit for these notes. This work was supported in part 
by the National Science Foundation.



\section*{References}

\begin{thebibliography}{999}

\bibitem{BBNS99}
M. Beneke, G. Buchalla, M. Neubert and C.T. Sachrajda,
\prl{83}{1999}{1914}.

\bibitem{BBNS00}
M. Beneke, G. Buchalla, M. Neubert and C.T. Sachrajda,
\npb{591}{2000}{313}.

\bibitem{IW89}
N. Isgur and M.B. Wise, \plb{232}{1989}{113};
\ibid{237}{1990}{527}.

\bibitem{VS87}
M.A. Shifman and M.B. Voloshin, \sjnp{45}{1987}{292} 
[\yf{45}{1987}{463}];
\ibid{47}{1988}{511} [{\bf 47} (1988) 801].

\bibitem{BBL}
For a review, see: 
G. Buchalla, A.J. Buras and M.E. Lautenbacher, \rmp{68}{1996}{1125}. 

\bibitem{FS78}
D. Fakirov and B. Stech, \npb{133}{1978}{315}.

\bibitem{CaMa78}
N. Cabibbo and L. Maiani, \plb{73}{1978}{418}; 
\ibid{76}{1978}{663} (E).

\bibitem{LB80}
G.P. Lepage and S.J. Brodsky, \prd{22}{1980}{2157}.

\bibitem{EfRa80}
A.V. Efremov and A.V. Radyushkin, \plb{94}{1980}{245}.

\bibitem{Bj89}
J.D. Bjorken, \npps{11}{1989}{325}.

\bibitem{DG91}
M.J. Dugan and B. Grinstein, \plb{255}{1991}{583}.

\bibitem{BBNSfuture}
M. Beneke, G. Buchalla, M. Neubert and C.T. Sachrajda,
{\it QCD factorization for $B\to\pi K$ decays}, 
Preprint \hepph{0007256}, to appear in the Proceedings of the 30th 
International Conference on High-Energy Physics (ICHEP 2000), Osaka, 
Japan, 27 July--2 August 2000, and paper in preparation.

\bibitem{MT90}
L. Maiani and M. Testa, \plb{245}{1990}{585}.

\bibitem{review}
For a review, see:
M. Neubert, \prep{245}{1994}{259}.

\bibitem{DGPS96}
J.F. Donoghue, E. Golowich, A.A. Petrov and J.M. Soares,
\prl{77}{1996}{2178}.

\bibitem{NeSt97}
For a review, see:
M. Neubert and B. Stech, in: {\it Heavy Flavours II}, ed.\ 
A.J.~Buras and M.~Lindner (World Scientific, Singapore, 1998) 
pp.~294 [hep-ph/9705292];\\
M. Neubert, \npps{64}{1998}{474}.

\bibitem{PW91}
H.D. Politzer and M.B. Wise, \plb{257}{1991}{399}.

\bibitem{Khod98a}
A. Khodjamirian and R. R\"uckl, in: {\it Heavy Flavours II}, ed.\ 
A.J.~Buras and M.~Lindner (World Scientific, Singapore, 1998) 
pp.~345 [hep-ph/9801443].

\bibitem{BF90}
V.M. Braun and I.E. Filyanov, \zpc{48}{1990}{239}.

\bibitem{GT82}
B.V. Geshkenbein and M.V. Terentev, \plb{117}{1982}{243};
\sjnp{39}{1984}{554} [\yf{39}{1984}{873}].

\bibitem{CLEO9701}
B. Barish et al., CLEO Collaboration, Conference report 
CLEO CONF~97-01 (EPS~97-339).

\bibitem{CLEOsupp}
B. Nemati et al., CLEO Collaboration, \prd{57}{1998}{5363}. 

\bibitem{Rodr97}
J.L. Rodriguez, in: Proceedings of the 2nd International Conference on 
{\it B Physics and CP Violation}, Honolulu, Hawaii, March 1997, ed.\ 
T.E. Browder et al.\ (World Scientific, Singapore, 1998) pp.~124 
[\hepex{9801028}].

\bibitem{PDG}
C. Caso et al., Particle Data Group, \epjc{3}{1998}{1}.

\bibitem{Lietal}
Y.Y. Keum, H.-N. Li and A.I. Sanda, 
{\it Fat penguins and imaginary penguins in perturbative QCD},
Preprint \hepph{0004004}.

\end{thebibliography}
\end{document}